\documentclass[authoryear,12pt]{elsarticle}

\newcommand{\bsigma}{\mbox{\boldmath $\sigma$}}
\def\nn{\nonumber}

\usepackage{graphics}
\usepackage{graphicx}
\usepackage{amssymb}
\usepackage{amsthm}
\usepackage{amsmath}
\journal{Physica E}

\begin{document}

\begin{frontmatter}


\title{Kohn Anomaly in Raman Spectroscopy of Single Wall Carbon Nanotubes}


\author{Ken-ichi Sasaki}
\address{International Center for Materials Nanoarchitectonics, 
National Institute for Materials Science, Namiki, Tsukuba 305-0044,
 Japan}

\author{Hootan Farhat}
\address{Department of Materials Science and Engineering,
Massachusetts Institute of Technology, Cambridge, MA 02139-4307}

\author{Riichiro Saito}
\address{Department of Physics, Tohoku University,Sendai 980-8578, Japan}

\author{Mildred S. Dresselhaus}
\address{Department of Physics, Department of Electrical Engineering
and Computer Science, Massachusetts Institute of Technology, Cambridge,
MA 02139-4307}

%\author{Katsunori Wakabayashi}
%\address{International Center for Materials Nanoarchitectonics, 
%National Institute for Materials Science,
%Namiki, Tsukuba 305-0044, Japan}
%\address{PRESTO, Japan Science and Technology Agency,
%Kawaguchi 332-0012, Japan}
%
%\author{Toshiaki Enoki}
%\address{Department of Chemistry, Tokyo Institute of Technology,
%Ookayama, Meguro-ku, Tokyo 152-8551, Japan} 

\begin{abstract}
 Phonon softening phenomena of the $\Gamma$ point optical modes 
 including the longitudinal optical mode, transverse optical mode 
 and radial breathing mode in ``metallic'' single wall carbon nanotubes
 are reviewed from a theoretical point of view. 
 The effect of the curvature-induced mini-energy gap on the phonon
 softening which depends on the Fermi energy and chirality of the
 nanotube is the main subject of this article. 
 We adopt an effective-mass model with a deformation-induced gauge
 field which provides us with a unified way to discuss the 
 curvature effect and the electron-phonon interaction.
\end{abstract}


\end{frontmatter}
%%\linenumbers


\section{Introduction}

The lattice structure of a single wall carbon nanotube 
(SWNT) can be specified uniquely
by the chirality defined by two integers $(n,m)$ 
[\cite{saito92apl,saito92prb}], 
and the chirality can be determined by Raman spectroscopy
[\cite{l818,jorio03,i1049}].
A simple tight-binding model shows that 
a SWNT is primarily metallic if $n-m$ is a multiple of 3 
or semiconducting otherwise.
A ``metallic'' SWNT can have a mini-energy band gap
due to the curvature of a SWNT which 
gives rise to a hybridization 
between the $\sigma$ and $\pi$ orbitals.
The presence of an energy band gap 
in a metallic SWNT has attracted much attention 
since the early stages of nanotube research 
[\cite{hamada92,mintmire92}]. 
The present paper deals with the effect of curvature 
on the Raman spectra for two in-plane 
$\Gamma$ point longitudinal and transverse 
optical phonon (LO and TO) modes 
[\cite{farhat07,sasaki08_curvat}] and 
the out-of-plane radial breathing mode (RBM)
[\cite{farhat09,sasaki08_chiral}].

In the Raman spectra of a SWNT, 
the LO and TO phonon modes at the $\Gamma$
point in the two-dimensional Brillouin zone (2D
BZ), which are degenerate in graphite and
graphene, split into two peaks, denoted by $G^+$ and $G^-$
peaks, respectively, [\cite{jorio03,saito98book,saito03}]
because of the curvature effect.
The splitting of the two
peaks for SWNTs is inversely proportional to the
square of the diameter, $d_t$, of SWNTs due to the
curvature effect, in which $G^+$ does not change
with changing $d_t$, but the $G^-$ frequency decreases
with decreasing $d_t$ [\cite{jorio02}].
In particular, for
metallic SWNTs, the $G^-$ peaks appear at a lower
frequency than the $G^-$ peaks for semiconducting
SWNTs with a similar diameter [\cite{w699}].
The spectra of $G^-$
for metallic SWNTs show a much larger spectral
width than that for semiconducting SWNTs.

It has been widely accepted 
that the frequency shift of the $G$-band 
in metallic SWNTs
is produced by the electron-phonon (el-ph)
interaction
[\cite{piscanec04,lazzeri06prl,ishikawa06,popov06,caudal07,das07}].
An optical phonon 
changes into an electron-hole pair
as an intermediate state 
by the el-ph interaction. 
This process is responsible for the phonon self-energy. 
The phonon self-energy is 
sensitive to the structure of the Fermi surface [\cite{kohn59}]
or the Fermi energy, $E_{\rm F}$.
In the case of graphite intercalation compounds 
in which the charge transfer of an electron 
from a dopant to the graphite layer 
can be controlled by the doping atom and its concentration, 
\cite{eklund77} observed 
a shift of the $G$-band frequency 
with an increase of the spectral width.
In this case the frequency shifted spectra show that 
not only the LO mode but also the TO mode
is shifted in the same fashion by a dopant.
For a graphene mono-layer,
Lazzeri {\it et al.} calculated 
the $E_{\rm F}$ dependence of the shift 
of the $G$-band frequency [\cite{lazzeri06prl}].
The LO mode softening in metallic SWNTs was shown by
~\cite{dubay02,dubay03} 
on the basis of density functional theory.
Recently \cite{nguyen07} 
and \cite{farhat07}
observed the phonon softening effect of SWNTs experimentally
as a function of $E_{\rm F}$ by electro-chemical doping,
and their results clearly show that the LO
phonon modes become soft as a function of $E_{\rm F}$.
\cite{ando08} discussed the phonon softening for metallic SWNTs
as a function of the $E_{\rm F}$ position, 
in which the phonon softening occurs for the LO phonon mode.
In this paper, we consider the effect of a curvature-induced mini-energy
gap on the frequency of the LO, TO, and RBM in ``metallic'' SWNTs.

The organization of the paper is as follows. 
In Sec.~\ref{sec:curv}
we show that the curvature of a SWNT gives rise to 
a hybridization between the $\sigma$ and $\pi$ orbitals.
Then we show our calculated result for the curvature-induced mini-energy
gap appearing in ``metallic'' SWNTs.
The current status of the scanning tunneling spectroscopy experimental
results is briefly mentioned,
confirming the curvature-induced mini-energy gap.
In Sec.~\ref{sec:curv-phonon}
we formulate the phonon self-energy
which is given by the electron-hole pair creation process.
The Fermi energy dependence of the self-energy
is shown for graphene with or without an energy gap, as a simple example.
In Sec.~\ref{sec:elph} 
we provide a theoretical framework 
for including a lattice deformation
into an effective-mass Hamiltonian.
A lattice deformation is represented 
by a deformation-induced gauge field
which is shown to be a useful idea 
to discuss both the appearance of the 
curvature-induced mini-energy gap
and also the el-ph interaction.
Sec.~\ref{sec:main} is a main section 
in this article in which we discuss the 
effect of curvature on the phonon self energy.
In Sec.~\ref{sec:dis} 
we discuss and summarize our results.

\section{Curvature Effect}
\label{sec:curv}

Let us start to discuss
the effect of the curvature of a SWNT on the hybridization 
between the $\sigma$ and $\pi$ orbitals (Sec.~\ref{ssec:cih}), 
and we then show the calculated result of the 
curvature-induced mini-energy gap 
appearing in ``metallic'' SWNTs (Sec.~\ref{ssec:gap}).
The phonon softening phenomena are sensitive to this mini-energy gap.

\subsection{Curvature-Induced Hybridization}\label{ssec:cih}

\begin{figure}[htbp]
 \begin{center}
  \includegraphics[scale=0.7]{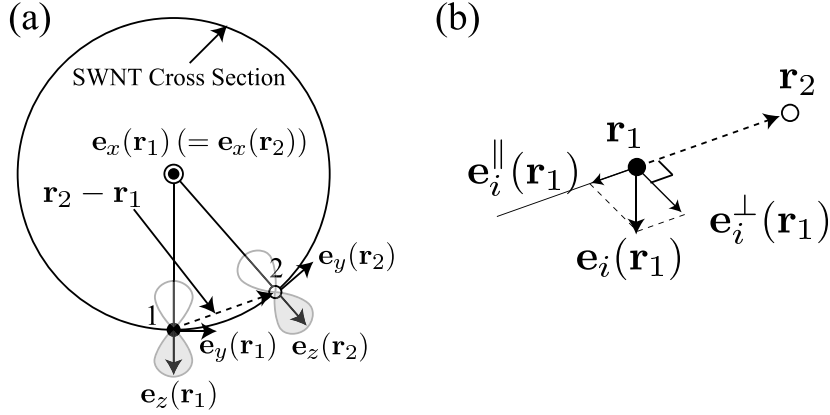}
 \end{center}
 \caption{
 (a) The curvature-induced hybridization between two $p_z$ orbitals 
 of carbon atoms at ${\bf r}_1$ and ${\bf r}_2$ is illustrated. 
 ${\bf e}_i({\bf r})$ ($i\in \{x,y,z\}$) denotes the basis of the 
 $(x,y,z)$-coordinate system whose origin is located at a carbon atom ${\bf r}$.
 (b) ${\bf e}_i^\parallel({\bf r}_2)$ and 
 ${\bf e}_i^\parallel({\bf r}_1)$ induce the hybridization, including
 $\sigma$ bonding.
 }
 \label{fig:curvature}
\end{figure}

At each carbon atom located at ${\bf r}$ 
on the surface of a SWNT,
we define the atom-specific $(x,y,z)$-coordinate axes and
the unit vector for each axis 
by ${\bf e}_{i}({\bf r})$ ($i\in \{x,y,z\}$), where
${\bf e}_{z}({\bf r})$ is taken as the unit normal vector to the cylindrical surface,
and ${\bf e}_{x}({\bf r})$ and ${\bf e}_{y}({\bf r})$ are unit vectors 
in the tangent plane [see Fig.~\ref{fig:curvature}(a)].
Here, 
${\bf e}_{x}({\bf r})$ is taken to be parallel to the axis of a SWNT.
In the case of a flat graphene sheet,
we can set the common axis vector for all carbon atoms and thus
a unit vector ${\bf e}_{i}$ at ${\bf r}_1$ can be taken orthogonal 
to the other ${\bf e}_{j}$ at ${\bf r}_2$ so that 
${\bf e}_{i}({\bf r}_1) \cdot {\bf e}_{j}({\bf r}_2) = \delta_{ij}$.
For SWNTs, however the orthogonal conditions are not satisfied because
of the atom specific coordinate, that is,
${\bf e}_{z}({\bf r}_1) \cdot {\bf e}_{z}({\bf r}_2)\ne1$,
${\bf e}_{z}({\bf r}_1) \cdot {\bf e}_{y}({\bf r}_2)\ne0$,
etc.

To see the curvature effect more clearly, 
it is useful to project ${\bf e}_{i}({\bf r}_1)$
and ${\bf e}_{j}({\bf r}_2)$ into 
\begin{align}
 {\bf e}_{i}({\bf r}_1) =
 {\bf e}_{i}^\perp({\bf r}_1)+ {\bf e}_{i}^\parallel({\bf r}_1),
 \ \
 {\bf e}_{j}({\bf r}_2) =
 {\bf e}_{j}^\perp({\bf r}_2)+ {\bf e}_{j}^\parallel({\bf r}_2),
 \label{eq:e_para_perp}
\end{align}
where $\parallel$ ($\perp$) denotes the vector which is parallel
(perpendicular) to the displacement vector ${\bf r}_2-{\bf r}_1$
[see Fig.~\ref{fig:curvature}(b)].
Let $| p_{i}({\bf r}) \rangle$ ($i\in \{x,y,z\}$) be the $2p_i$-orbital
of a carbon atom located at ${\bf r}$.
Then, the transfer integral
from $| p_{i}({\bf r}_1) \rangle$ to $|p_{j}({\bf r}_2) \rangle$ 
may be written as
\begin{align}
 \langle p_{j}({\bf r}_2) | {\hat {\cal H}} | p_{i}({\bf r}_1) \rangle 
 = {\cal H}_{pp\pi}
 {\bf e}_{j}^\perp({\bf r}_2) \cdot {\bf e}_{i}^\perp({\bf r}_1)+
 {\cal H}_{pp\sigma}
 {\bf e}_{j}^\parallel({\bf r}_2) \cdot {\bf e}_{i}^\parallel({\bf r}_1),
 \label{eq:trans1}
\end{align}
where ${\cal H}_{pp\pi}$ and ${\cal H}_{pp\sigma}$ 
are the transfer integrals for $\pi$ and $\sigma$ bonds,
respectively.
According to a first-principles calculation 
with the local density approximation obtained by~\cite{porezag95},
${\cal H}_{pp\pi}\approx -3$ eV and ${\cal H}_{pp\sigma}\approx 8$ eV
for nearest-neighbor carbon sites.
Using Eq.~(\ref{eq:e_para_perp}),
we eliminate ${\bf e}_i^\perp({\bf r}_1)$ and 
${\bf e}_j^\perp({\bf r}_2)$ from Eq.~(\ref{eq:trans1}),
and get
\begin{align}
 \langle p_{j}({\bf r}_2)  | {\hat {\cal H}} | p_{i}({\bf r}_1) \rangle 
 = {\cal H}_{pp\pi} {\bf e}_j({\bf r}_2)  \cdot {\bf e}_i({\bf r}_1) 
 + \left( {\cal H}_{pp\sigma} - {\cal H}_{pp\pi} \right)
 {\bf e}_j^\parallel({\bf r}_2)  \cdot {\bf e}_i^\parallel({\bf r}_1),
 \label{eq:trans2}
\end{align}
where we have used ${\bf e}_i^\perp({\bf r}_1) \cdot {\bf e}_j^\parallel({\bf r}_2)=0$
and ${\bf e}_i^\parallel({\bf r}_1) \cdot {\bf e}_j^\perp({\bf r}_2)=0$.
The last term of Eq.~(\ref{eq:trans2}) corresponds to 
the curvature effect of a SWNT.
Note that the coefficient of the last term includes 
${\cal H}_{pp\sigma}$ showing that the
$\sigma$ bond is partially incorporated by 
the curvature-induced hybridization [See~\cite{ando00} for more details].

%%%%%%%%%%%%%%%%%%%%%%%%%%%%%
%%%%%%%%%%%%%%%%%%%%%%%%%%%%%

In the case of a flat graphene, we have 
${\bf e}_z^\parallel({\bf r}_1)=0$ and ${\bf e}_z^\parallel({\bf r}_2)=0$.
Then, the last term of Eq.~(\ref{eq:trans2}) disappears
and the theoretical model taking only 
the $2p_z$ orbital (or $\pi$-orbital) into account
becomes a good approximation.
The curvature of a SWNT results in 
${\bf e}_z^\parallel({\bf r}_2)\cdot{\bf e}_z^\parallel({\bf r}_1) \ne
0$,
${\bf e}_z^\parallel({\bf r}_2)\cdot{\bf e}_y^\parallel({\bf r}_1) \ne
0$, etc., and the last term of Eq.~(\ref{eq:trans2})
is non-vanishing and consequently 
the curvature-induced hybridization occurs.
The curvature-induced hybridization
is relevant to the following two physical properties.
First, the hybridization can open a mini-gap 
(up to $\sim$ 100meV) near the Fermi energy in metallic SWNTs. 
Second, the curvature-induced gap depends on the SWNT $(n,m)$ chirality.
For example, the gap is zero for armchair SWNTs, 
while it is about 70 meV for a $(12,0)$ metallic zigzag SWNT.
The chirality dependent curvature-induced energy gap 
will be analytically given in the next subsection.

\subsection{Curvature-Induced Mini-Energy Gap}\label{ssec:gap}

\begin{figure}[htbp]
 \begin{center}
  \includegraphics[scale=0.8]{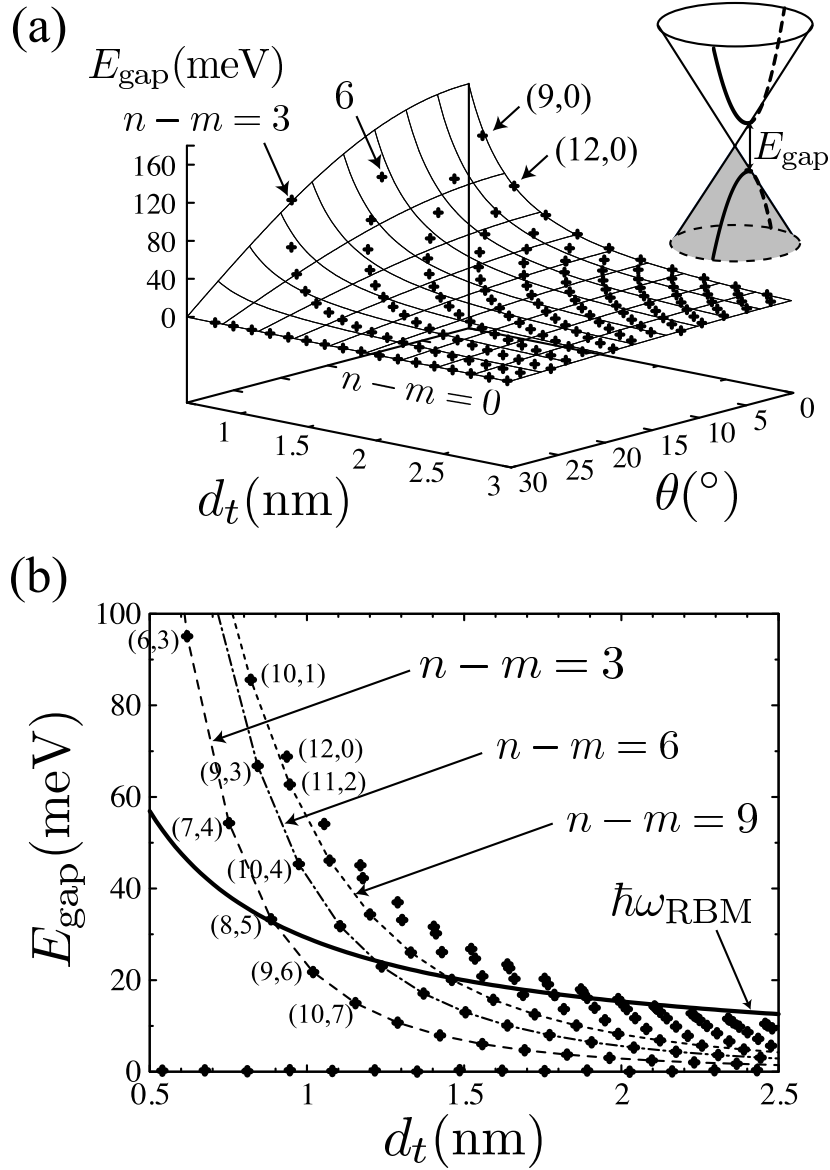}
 \end{center}
 \caption{
 (a)
 The dependence of the curvature-induced energy gap, $E_{\rm gap}$,
 on the chiral angle $\theta$ and tube diameter $d_t$.
 The surface is a plot of Eq.~(\ref{eq:Egap}) 
 which reproduces well the calculated results. 
 (inset) Due to the nanotube curvature, the cutting line 
 which was exactly crossing the Dirac point 
 in the absence of curvature can miss the Dirac point
 when curvature is included.
 This curvature gives rise to an energy gap $E_{\rm gap}$
 in ``metallic'' SWNTs.
 (b) The $d_t$ dependence of $E_{\rm gap}$
 is given as a one-dimensional projection of (a)
 onto the $d_t$ axis.
 The points on the dashed, dot-dashed, and dotted curves 
 satisfy $n-m=3,6,9$, respectively.
 We plot the energy of the RBM, $\hbar \omega_{\rm RBM}$ 
 of Eq.~(\ref{eq:rbm_ene}), as a solid curve for comparison.
 }
 \label{fig:Erbm}
\end{figure}

In Fig.~\ref{fig:Erbm}(a) 
we plot the calculated curvature-induced energy gap, $E_{\rm gap}$, 
for each $(n,m)$ for metallic SWNTs
as a function of the chiral angle $\theta(^\circ)$ 
and tube diameter $d_t$(nm).
We performed the energy band structure calculation 
in an extended tight-binding (ETB) framework developed 
by~\cite{samsonidze04} to obtain $E_{\rm gap}$.
In the ETB framework,
$2s$ and $2p$ orbitals, 
and their transfer and overlap integrals up to fourth nearest
neighbor atoms are taken into account
[see~\cite{popov04,samsonidze04} for more details].~\footnote{In the ETB
program, we numerically solve the energy eigenequation,  
${\hat {\cal H}}|\Psi\rangle =E|\Psi\rangle$, 
in the basis of $|s({\bf r}) \rangle$ and $|p_{i}({\bf
r})\rangle$ for two carbon atoms (A and B). 
The basis orbitals for the A-atom are non-orthogonal to 
those for the B-atom due to the curvature effect, and 
the Hamiltonian and overlap matrices are $8\times 8$ matrices.
We assumed the on-site energies
$E(2p)=-4.882$[eV] and
$E(2s)=-13.573$[eV].
$E(2s)-E(2p)\approx -8.7$[eV] is
close to the value ($-8.868$[eV]) shown in \cite{saito92prb}.}
We have adopted the values of the transfer and overlap integrals
as a function of the carbon-carbon inter-atomic distance
that were derived by~\cite{porezag95}.~\footnote{
Although the energy gap at the Fermi level has little to do with the
overlap integral, we shall note that the overlap
integrals $S^{\rm CC}_{pp\sigma}$ and $S^{\rm CC}_{pp\pi}$ are switched
in Table I of \cite{porezag95}.}

Figure~\ref{fig:Erbm}(a) shows that, 
for a fixed diameter of a metallic SWNT
$d_t$, a zigzag SWNT ($\theta=0^\circ$)
has the largest value of $E_{\rm gap}$
and an armchair SWNT ($\theta=30^\circ$) has no energy gap.
%The $(n,m)$ values associated with the curves are given by
%$n-m=3q$ ($q=0,1,2,\cdots$).
The calculated results are well reproduced by
\begin{align}
 E_{\rm gap} = \frac{c}{d_t^2} \cos 3\theta,
 \label{eq:Egap}
\end{align}
with $c=60$(eV$\cdot$nm$^2$) [\cite{sasaki08_chiral}].
The chirality and diameter dependence of $E_{\rm gap}$
is consistent with the results by~\cite{kane97}, and~\cite{ando00}.
The value of $c$ is about two times larger than the result by~\cite{kane97}. 
This difference may come from the inclusion of ${\cal H}_{pp\sigma}$ 
in our calculation.
As we will explain in detail in Sec.~\ref{subsec:digf}, 
the curvature moves the Dirac point in $k$-space
away from the hexagonal corner of the first BZ.
As a result, the curvature can cause 
the quantized transverse electron wave vector 
(the cutting line) to miss the Dirac point
and make a gap [see the inset in Fig.~\ref{fig:Erbm}(a)].

%%%%%%%%%%%%%%%%%%%%%%%%%%%%%
%%%%%%%%%%%%%%%%%%%%%%%%%%%%%

When we discuss the phonon softening of the RBM,
the relationship between the mini-energy gap
and the RBM phonon energy will be important.
In Fig.~\ref{fig:Erbm}(b),
we plot the energy of the RBM,
\begin{align}
 \hbar \omega_{\rm RBM} = \frac{c_1}{d_t} + c_2,
 \label{eq:rbm_ene}
\end{align}
as a solid curve for comparison.
Here $\hbar \omega_{\rm RBM}$ is a monotonic function 
of the tube diameter ($d_t$[nm])
and is modeled as being linear in the inverse diameter, with an offset
$c_2$ which is known as the effect of the substrate.
We assume that $c_1=223.5$[cm$^{-1}$] and $c_2=12.5$[cm$^{-1}$]
which are experimentally derived parameters as obtained 
by~\cite{strano03} and \cite{bachilo02}.
Using Eqs.~(\ref{eq:rbm_ene}) and (\ref{eq:Egap}) 
for zigzag SWNTs ($\theta=0^\circ$),
we see that $E_{\rm gap}$ is smaller than $\hbar \omega_{\rm RBM}$
when $d_t > 2$[nm] (see Fig.~\ref{fig:Erbm}(b)).

The presence (absence) of a curvature-induced mini-energy gap 
in ``metallic'' zigzag (armchair) SWNTs
was confirmed experimentally by~\cite{ouyang01}.
The chirality was measured experimentally for $(9,0)$, $(12,0)$, and
$(15,0)$ zigzag SWNTs by these authors.
The observed energy gap can be fitted by $4A_0/d_t^2$ 
which has the same $d_t$ dependence in Eq.~(\ref{eq:Egap}).
Note that the coefficient is given by $A_0=3\gamma_0 a_{\rm cc}^2/16$, 
and $4A_0\approx 40$[meV$\cdot$nm$^2$] 
is smaller than the value of $c=60$[meV$\cdot$nm$^2$] 
in Eq.~(\ref{eq:Egap}).~\footnote{
Putting $\gamma_0=2.60$[eV] and $a_{\rm cc}=0.142$[nm]
into the definition of $A_0$, 
we get the result $4A_0\approx 40$[meV$\cdot$nm$^2$].}
This discrepancy may be attributed to 
(1) uniaxial and torsional strain 
which is unintentionally applied to a
SWNT [\cite{yang99,yang00,kleiner01}],~\footnote{
We expect that the curvature-induced gap follows 
(see Sec.~\ref{subsec:digf} for the derivation)
\begin{align}
 E_{\rm gap} = \left\{ \frac{c}{d_t^2} -a \right\} \cos 3\theta,
\end{align}
when an uniaxial strain is applied to SWNTs. The value of $a$ depends on
the model used, but it is probably not dependent on $d_t$. Considering
the fact that the observed energy gap scales as $d_t^{-2}$, the effect
of strain is not so relevant.
} or 
(2) renormalization of the value of $c$ due to the el-ph interaction, or
(3) a SWNT-substrate interaction effect.
(1,2) are intrinsic to SWNTs, while (3) is extrinsic.
Since there are various factors 
which can affect the energy gap, 
it is not easy to predict the precise value of the energy gap,
although the curvature-induced gap
has been examined within the framework of first principles calculations
including the effect of structure optimization [\cite{miyake05}].
It is noted that 
the chirality dependence of $\cos 3\theta$ 
in Eq.~(\ref{eq:Egap}) has not been tested experimentally 
so far, except for $\theta=0$ (zigzag SWNTs) 
and $\theta=30^\circ$ (armchair SWNTs).
Study of a chiral SWNT is left for future experiments.

\section{Effect of Curvature on the Phonon Energy}
\label{sec:curv-phonon}

In this section 
we formulate the self-energy of a phonon mode (Sec.~\ref{ssec:selfene}),
and explain qualitatively the effect of the curvature on the self-energy
(Sec.~\ref{ssec:softhard}). 
The relationship between our formulation and that of others is referred
to in Sec.~\ref{ssec:others}.

\subsection{Phonon Self-Energy}\label{ssec:selfene}

A renormalized phonon energy is written 
as a sum of the unrenormalized energy, $\hbar \omega$,
and the real part of the self-energy, $\Pi(\omega,E_{\rm F})$.
The imaginary part of $\Pi(\omega,E_{\rm F})$ gives 
the spectrum width.
Throughout this paper, 
we assume a constant value for $\hbar \omega$
for each phonon mode.
The self-energy is given by
time-dependent second-order perturbation theory as
\begin{eqnarray}
 \Pi(\omega,E_{\rm F}) = 
 2 \sum_{\bf k} \left(
 \frac{|V_{\bf k}|^2}{\hbar \omega -E^{\rm eh}_{\bf k}+i\Gamma/2}
 - \frac{|V_{\bf k}|^2}{\hbar \omega +E^{\rm eh}_{\bf k}+i\Gamma/2}
 \right)
 \times \left(f_{\rm h}-f_{\rm e}\right),
 \label{eq:PI}
\end{eqnarray}
where the pre-factor 2 comes from spin degeneracy,
$f_{\rm h,e}=(1+\exp(\beta(E^{\rm h,e}-E_{\rm F}))^{-1}$
is the Fermi distribution function,
$E^{\rm e}_{\bf k}$ ($E^{\rm h}_{\bf k}$) 
is the energy of an electron (a hole) with momentum ${\bf k}$,
and $E^{\rm eh}_{\bf k} \equiv E^{\rm e}_{\bf k}-E^{\rm h}_{\bf k}$ ($\ge 0$)
is the energy of an electron-hole pair. 
$V_{\bf k}$ is the el-ph matrix element that 
a phonon with momentum ${\bf q}=0$
changes into an electron-hole pair 
[see the left diagram of Fig.~\ref{fig:2ndpert}(a)]
which will be derived in Sec.~\ref{sec:elph}.
Note that the momentum of an electron ${\bf k}$ is the same as that of
a hole due to momentum conservation, and therefore pair creation 
involves a vertical transition. 
In Eq.~(\ref{eq:PI}), 
the energy shift is given by the real part of the self-energy,
${\rm Re}[\Pi(\omega,E_{\rm F})]$, and 
the decay width $\Gamma$ is determined self-consistently
by $\Gamma/2= - {\rm Im} \left[ \Pi(\omega,E_{\rm F})
\right]$.~\footnote{The self-consistent calculation begins by putting 
$\Gamma/2=\gamma_0$ into the right-hand side of Eq.~(\ref{eq:PI}).
By summing the right-hand side, we have a new $\Gamma/2$ via
$\Gamma/2= - {\rm Im} \left[ \Pi(\omega,E_{\rm F}) \right]$ and 
we then put the new $\Gamma/2$ into the right-hand side again,
iteratively. 
This calculation is repeated until $\Pi(\omega,E_{\rm F})$ is
converged.}
The decay width relates to the average life-time $\tau$ via
$\tau = \hbar/\Gamma$.
It is noted that we use $T=300$K 
although the self-energy is also a function of temperature 
[$\beta^{-1}=k_{\rm B}T$ where $k_{\rm B}$ is Boltzmann's constant].

\subsection{Phonon Softening and Hardening}\label{ssec:softhard}

\begin{figure}[htbp]
 \begin{center}
  \includegraphics[scale=0.7]{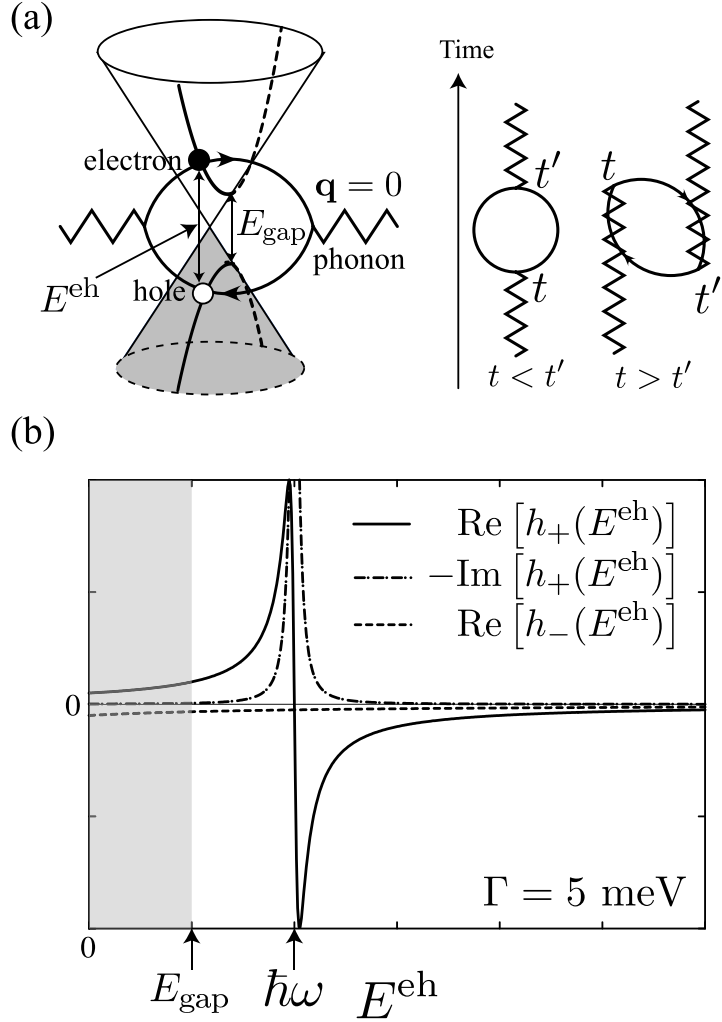}
 \end{center}
 \caption{
 (a) In time-dependent second-order perturbation theory,
 we consider an intermediate state including only electron-hole pairs
 (the case of $t<t'$), and an intermediate state including two phonons
 and electron-hole pairs (the case of $t>t'$).
 The former process corresponds to $h_{+}(E^{\rm eh})$, while the latter
 one corresponds to $h_{-}(E^{\rm eh})$.
 (b) The energy correction to the phonon energy 
 by an intermediate electron-hole pair,
 especially the sign of ${\rm Re}(h_{+}(E^{\rm eh}))$ (solid curve),
 that corresponds to frequency hardening or softening,
 depends on the energy of the intermediate state $E^{\rm eh}$.
 The contribution to $\Pi(\omega,E_{\rm F})$
 of a low energy electron-hole pair 
 satisfying $0 \le E^{\rm eh} \le E_{\rm gap}$ is forbidden.
 ${\rm Im}(h_{+}(E^{\rm eh}))$ (dashed curve)
 is nonzero only when $E^{\rm eh}$ is very close to $\hbar \omega$,
 which shows that a phonon mode can resonantly 
 decay into an electron-hole pair 
 with the same energy.
 }
 \label{fig:2ndpert}
\end{figure}

By defining the denominators of Eq.~(\ref{eq:PI}) as 
$h_{\pm}(E^{\rm eh})\equiv \pm1/(\hbar \omega \mp E^{\rm eh}+i\Gamma/2)$, 
Eq.~(\ref{eq:PI}) may be rewritten as
\begin{align}
 \Pi(\omega,E_{\rm F}) = 
 2 \sum_{\bf k} |V_{\bf k}|^2 \left[
 h_{+}(E^{\rm eh}_{\bf k}) + h_{-}(E^{\rm eh}_{\bf k}) 
 \right]
 \times \left(f_{\rm h}-f_{\rm e}\right).
 \label{eq:PI2}
\end{align}
When we assume that $|V_{\bf k}|^2$ does not depend on ${\bf k}$, 
the $E^{\rm eh}$ dependence of ${\rm Re}[\Pi(\omega,E^{\rm eh})]$
is determined by those of ${\rm Re}[h_{+}(E^{\rm eh})]$ and ${\rm Re}[h_{-}(E^{\rm eh})]$.
It should be noted that ${\rm Re}[h_{+}(E^{\rm eh})]$ 
(solid curve in Fig.~\ref{fig:2ndpert}(b))
has a positive (negative) value
when $E^{\rm eh} < \hbar \omega$ ($E^{\rm eh} > \hbar \omega$), and
the lower (higher) energy electron-hole pair
makes a positive (negative) contribution to 
${\rm Re}[\Pi(\omega,E_{\rm F})]$.
Therefore, 
the sign of the contribution to ${\rm Re} [\Pi(\omega,E_{\rm F})]$,
i.e., frequency hardening or softening, depends on its electron-hole
virtual state energy, $E^{\rm eh}$.
In contrast, ${\rm Re}[h_{-}(E^{\rm eh})]$ 
(dashed curve in Fig.~\ref{fig:2ndpert}(b)) always has a negative value,
that is, it only contributes to a phonon softening. 
Note however that the contribution of ${\rm Re}[h_{-}(E^{\rm eh})]$ 
is small compared with ${\rm Re}[h_{+}(E^{\rm eh})]$ 
since $-1/\hbar \omega \le {\rm Re}[h_{-}(E^{\rm eh})] < 0$.
Physically speaking, the $h_{-}(E^{\rm eh})$ term represents 
an intermediate state including two phonons and electron-hole pairs
(see the right hand diagram in Fig.~\ref{fig:2ndpert}(a)), while 
the $h_{+}(E^{\rm eh})$ term represents 
the intermediate state that includes 
only electron-hole pairs.~\footnote{In
fact, we have $h_{-}(E^{\rm eh})=h_{+}(2(\hbar \omega+i\Gamma/2) +
E^{\rm eh})$.}
Even though the contribution of $h_{-}(E^{\rm eh})$
is relatively small, $h_{-}(E^{\rm eh})$ is important
to get a symmetric response of $\Pi(\omega,E_{\rm F})$
relative to the Fermi energy. In fact,
due to the $h_{-}(E^{\rm eh})$ term,
the electron-hole pair at the Dirac point 
($E^{\rm eh}=0$) can not contribute to the self-energy, since 
${\rm Re}[h_{+}(E^{\rm eh})+h_{-}(E^{\rm eh})]=0$ when $E^{\rm eh}=0$.
For high energy electron-hole pairs, 
the $h_{\pm}(E^{\rm eh})$ terms contribute equally since 
${\rm Re}[h_{+}(E^{\rm eh})]\approx {\rm Re}[h_{-}(E^{\rm eh})] \approx
-1/E^{\rm eh}$.

%%%%%%%%%%%%%%%%%%%%%%%%%%%%%
%%%%%%%%%%%%%%%%%%%%%%%%%%%%%

The curvature-induced energy gap, $E_{\rm gap}$, 
affects the frequency shift
since an electron-hole pair creation event is possible 
only when $E^{\rm eh} \ge E_{\rm gap}$.
When $0 < E_{\rm gap} \le \hbar \omega$,
the contribution to frequency hardening in Eq.~(\ref{eq:PI})
is suppressed.
When $E_{\rm gap} > \hbar \omega$,
not only are all the positive contributions to the self-energy
suppressed, 
but some negative contributions are also suppressed.
Further,
${\rm Im}(h_{+}(E^{\rm eh}))$ is nonzero 
only when $E^{\rm eh}$ is very close to 
$\hbar \omega$,
which shows that a phonon can resonantly decay into an electron-hole
pair with the same energy.
Thus, when $E_{\rm gap} > \hbar \omega$, we have 
$\Gamma \simeq 0$ 
because no resonant electron-hole pair excitation is allowed
near $E=\hbar \omega$.
It is therefore important to compare 
the values of $E_{\rm gap}$ and $\hbar \omega$
for each $(n,m)$ SWNT.
For the LO and TO modes, $\hbar \omega$ is about 0.2[eV] and therefore
we get $E_{\rm gap}< \hbar \omega$ (see Fig.~\ref{fig:Erbm}) for most of
the SWNTs except for a SWNT with a small diameter.~\footnote{For very
small diameter SWNTs, the energy gap disappears because of the lowering
of the interlayer energy bonds.}
Thus, those LO and TO modes can resonantly decay into an electron-hole
pair.  
The RBM mode in some SWNTs (for example, a $(12,0)$ zigzag SWNT)
can not resonantly decay into an electron-hole pair, which 
results in a long life-time for the RBM 
in that particular SWNT [\cite{sasaki08_chiral}].

At $T=0$, the Fermi distribution factor, 
namely $f_{\rm h}-f_{\rm e}$ in Eq.~(\ref{eq:PI}), 
plays a very similar role as the curvature-induced gap, $E_{\rm gap}$.
In fact, all the excitations of electron-hole pairs with $E^{\rm eh}\le 2|E_{\rm F}|$
are forbidden due to the Pauli exclusion principle. 
A difference between the energy gap and the Fermi energy arises at a
finite temperature.
Some electron-hole pairs with 
$E_{\rm gap} \le E^{\rm eh}\le 2|E_{\rm F}|$
can contribute to the self-energy, while 
states $E^{\rm eh}< E_{\rm gap}$ do not exist even 
at a finite temperature. 
It should be noted that $V_{\bf k}$ in Eq.~(\ref{eq:PI}) 
depends on the value of $E_{\rm gap}$ since the position of the cutting
line depends on $E_{\rm gap}$, while $V_{\bf k}$
does not change by changing $E_{\rm F}$. 
This is also a crucial difference between the roles of $E_{\rm gap}$ and
$E_{\rm F}$ in the self-energy.

\subsection{Other Formulas}\label{ssec:others}

Here, we refer to the relationship 
between our formula and other formulas.
First, replacing $\Gamma/2$ in Eq.~(\ref{eq:PI})
with a positive infinitesimal $0_+$ gives the standard formula
for the Fermi Golden rule.
In this case, using $1/(x+i0_+) = {\rm P}(1/x)-i\pi\delta(x)$
with ${\rm P}$ denoting the principle value of integration
and $\delta(x)$ the Dirac delta-function, 
$\Gamma/2$ can be calculated directly, i.e., 
without using the self-consistent way, by performing 
the summation (or integral) of the right-hand side of Eq.~(\ref{eq:PI}).
We calculate $\Gamma/2$ self-consistently by taking care of a finite
energy level spacing originating from a finite length of a nanotube
where $E^{\rm eh}_{\bf k}$ now takes a discrete value, 
and is not a continuous variable.
Roughly speaking, the broadening is suppressed when the energy level
spacing, $\Delta E = 2\pi \hbar v_{\rm F}/L$, exceeds $\Gamma$. 
For example, the critical length where the broadening becomes negligible
for a $(10,10)$ SWNT is about $700$nm.

Second, the summation index $\sum_{\bf k}$ in Eq.~(\ref{eq:PI}) 
is not restricted to only inter-band ($E^{\rm eh} \ne 0$) processes
but includes also intra-band ($E^{\rm eh}= 0$) processes.~\footnote{It
may be appropriate to denote an intra-band process by an
$E^{\rm ee}=0$ or $E^{\rm hh}=0$ process.}
Then, the self-energy can be decomposed into two parts, as 
$\Pi(\omega,E_{\rm F})=\Pi^{\rm inter}(\omega,E_{\rm F})
+\Pi^{\rm intra}(\omega,E_{\rm F})$ where
$\Pi^{\rm inter}(\omega,E_{\rm F})$ includes only inter-band processes 
satisfying $E^{\rm eh}\ne 0$.
In the adiabatic limit, 
i.e., when $\omega = 0$ and $\Gamma=0$ in Eq.~(\ref{eq:PI}),
it is straightforward to get the following relations, 
for a single Dirac cone at $T=0$:
\begin{align}
\begin{split}
 & \Pi^{\rm intra}(0,E_{\rm F})
 =-2\sum_{\bf k}|V_{\bf k}|^2 \times
 \begin{cases}
  f'(E^{\rm e}_{\bf k}), (E_{\rm F}>0) \\
  f'(E^{\rm h}_{\bf k}), (E_{\rm F}<0)
 \end{cases}
 = -\frac{\alpha}{2}|E_{\rm F}|, \\
 & \Pi^{\rm inter}(0,E_{\rm F})=
 -4\sum_{\bf k} \frac{|V_{\bf k}|^2}{E^{\rm eh}_{\bf k}}
 =-\frac{\alpha}{2} \left( E_c - |E_{\rm F}| \right),
\end{split}
\end{align}
where $\alpha\equiv S|v|^2/\pi(\hbar v_{\rm F})^2$
and $E_c$ is some cut-off energy.
Here, we have assumed that 
$V_{\bf k}=v\cos\Theta({\bf k})$.~\footnote{$\Theta({\bf k})$ is the
angle between the vector ${\bf k}$ and the $k_x$-axis [see Eq.~(\ref{eq:wfc})].}
Note that $\Pi^{\rm intra}(0,E_{\rm F})$ does not vanish 
because $(f_{\rm h}-f_{\rm e})/E^{\rm eh}_{\bf k} \ne 0$ in this limit, while
in the non-adiabatic case, 
$\Pi^{\rm intra}(\omega,E_{\rm F})$ 
vanishes since $(f_{\rm h}-f_{\rm e})/\hbar \omega=0$.
It is only the inter-band process that contributes to the self-energy 
in the non-adiabatic case.~\footnote{
\cite{lazzeri06prl} showed that $\Pi(0,E_{\rm F})$ 
does not depend on $E_{\rm F}$
in the adiabatic limit due to the cancellation
between $\Pi^{\rm intra}(0,E_{\rm F})$ and $\Pi^{\rm inter}(0,E_{\rm
F})$.
This shows that the adiabatic approximation is not appropriate for
discussing the $E_{\rm F}$ dependence of the self-energy.}
In the non-adiabatic limit at $T=0$, 
it is a straightforward calculation to get
\begin{align}
 {\rm Re} \left[ \Pi(\omega,E_{\rm F}) \right] = - \frac{\alpha}{2}
 \left[ E_c - |E_{\rm F}|- \frac{\hbar \omega}{4}
 \ln \left|  \frac{|E_{\rm F}| - \frac{\hbar \omega}{2}}{|E_{\rm F}|+
 \frac{\hbar \omega}{2} }\right| \right], \ \ 
 (E_c\gg \hbar \omega),
\end{align}
where $E^{\rm eh}_{\bf k}=2\hbar v_{\rm F}k$, 
$\sum_{\bf k}\to V/(2\pi)^2 \int_0^{k_c} kdk\int_0^{2\pi}d\Theta$,
and $\int x/(x+a) dx=x-a\ln|x+a|$ have been used 
in Eq.~(\ref{eq:PI}) to get the right-hand side.
The Fermi energy dependence is given by the last two terms 
for the case of a massless Dirac cone spectrum.
The first term is linear with respect to $E_{\rm F}$
and the second term produces a singularity at 
$|E_{\rm F}|=\hbar \omega/2$. 
This singularity is useful in identifying the actual Fermi energy 
of a graphene sample.

It is also interesting to consider the case of a massive Dirac cone
spectrum, $E=\pm\sqrt{m^2 + (\hbar v_{\rm F}k)^2}$. 
In the non-adiabatic limit at $T=0$, we get
\begin{align}
 {\rm Re} \left[ \Pi(\omega,E_{\rm F}) \right] = - \frac{\alpha}{2}
 \left[ E_c - |E_{\rm F}|- \left\{ \frac{ 
 \left( \hbar\omega \right)^2-(2m)^2}{4\hbar \omega} \right\}
 \ln \left|  \frac{|E_{\rm F}| - \frac{\hbar \omega}{2}}{|E_{\rm F}|+
 \frac{\hbar \omega}{2} }\right| \right],
 \label{eq:mass}
\end{align}
where $E_c\gg \hbar \omega$ and
$V_{\bf k}=v(\hbar v_{\rm F}k/E)\cos\Theta({\bf k})$ are assumed.
Equation~(\ref{eq:mass}) is for $|E_{\rm F}|\ge m$.
For $|E_{\rm F}|< m$, the self-energy shift is given by
replacing $|E_{\rm F}|$ with $m$ in Eq.~(\ref{eq:mass}).
The logarithmic singularity for the last term disappears
when $m = \hbar \omega/2$, and its overall sign is interchanged 
when $2m > \hbar \omega$. 
Broadening is possible only when $2m < \hbar \omega$, which may be
useful in knowing whether the graphene sample has an energy gap or not.

\section{The Electron-Phonon Interaction}\label{sec:elph}

In this section 
we provide a framework to obtain the el-ph (electron-phonon) interaction
in the effective-mass theory, and 
show how to calculate the el-ph matrix elements.
The main results are 
Eqs.~(\ref{eq:optigauge}) and (\ref{eq:elph_RBM_g}). 
Those who are not interested in the details 
of the derivation can skip this section.

\subsection{Unperturbed Hamiltonian}

\begin{figure}[htbp]
 \begin{center}
  \includegraphics[scale=1.0]{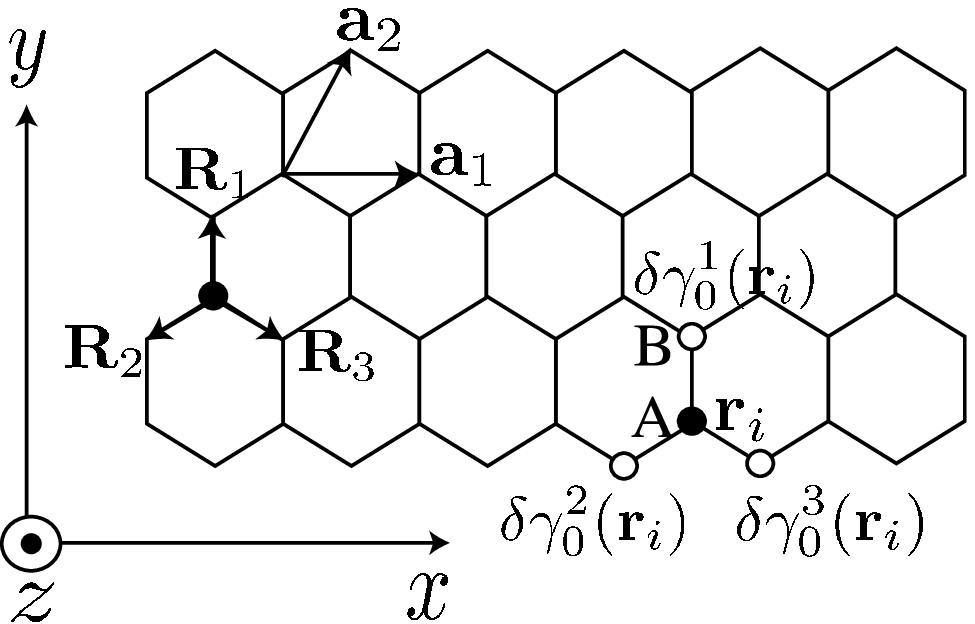}
 \end{center}
 \caption{A hexagonal unit cell of graphene consists of 
 {\rm A} (closed circle) and {\rm B} (open circle) sublattices.
 ${\bf a}_1$ and ${\bf a}_2$ are lattice vectors.
 ${\bf R}_a$ ($a=1,2,3$)
 are vectors pointing to the nearest-neighbor  
 {\rm B} sites from an {\rm A} site (${\bf R}_1=a_{\rm cc}{\bf e}_y$,
 ${\bf R}_2=-(\sqrt{3}/2)a_{\rm cc}{\bf e}_x -(1/2)a_{\rm cc}{\bf e}_y$,
 and 
 ${\bf R}_3=(\sqrt{3}/2)a_{\rm cc}{\bf e}_x -(1/2)a_{\rm cc}{\bf
 e}_y$) where
 ${\bf e}_x$ (${\bf e}_y$)
 is the dimensionless unit vector for the $x$-axis ($y$-axis).
 Local modulations of the hopping integral are defined by 
 $\delta \gamma^a_0({\bf r}_i)$ ($a=1,2,3$) where ${\bf r}_i$ 
 is the position of an A-atom.
 }
 \label{fig:graphene}
\end{figure}

The unperturbed Hamiltonian in the effective-mass model
for $\pi$-electrons near the K point of a graphene sheet 
is given by
\begin{align}
 {\cal H}^{\rm K}_0 = v_{\rm F} \bsigma \cdot \hat{\mathbf{p}},
 \label{eq:H0K}
\end{align}
where $v_{\rm F}$ is the Fermi velocity,
$\hat{\bf p}=-i\hbar \nabla$ is the momentum operator,
and $\bsigma=(\sigma_x,\sigma_y)$ is the Pauli matrix.~\footnote{
We use the Pauli matrices of the form of
$\sigma_x
=\begin{pmatrix}
  0 & 1 \cr 1 & 0
 \end{pmatrix}$, 
$\sigma_y
=\begin{pmatrix}
  0 & -i \cr i & 0
 \end{pmatrix}$, and
$\sigma_z
=\begin{pmatrix}
  1 & 0 \cr 0 & -1
 \end{pmatrix}$.
The $2 \times 2$ identity matrix $\sigma_0$ is given by
$\sigma_0 = \begin{pmatrix}
 1 & 0 \cr 0 & 1
\end{pmatrix}$.
}
The $x$, $y$, and $z$ coordinate system
is taken as shown in Fig.~\ref{fig:graphene}. 
${\cal H}^{\rm K}_0$ is a $2\times 2$ matrix 
which operates on the two component wavefunction:
\begin{align}
 \psi^{\rm K}({\bf r}) = 
 \begin{pmatrix}
  \psi^{\rm K}_{\rm A}({\bf r})\cr
  \psi^{\rm K}_{\rm B}({\bf r})
 \end{pmatrix},
\end{align}
where $\psi^{\rm K}_{\rm A}({\bf r})$ and $\psi^{\rm K}_{\rm B}({\bf r})$
are the wavefunctions of $\pi$-electrons
for the sublattices A and B, respectively, around the K point.
The energy eigenvalue of Eq.~(\ref{eq:H0K})
is given by $\pm v_{\rm F}|{\bf p}|$ and 
the energy dispersion relation
shows a linear dependence at the Fermi
point, which forms what is known as the Dirac cone.

The energy eigenstate with wave vector ${\bf k}$
in the conduction energy band is written by
a plane wave $e^{i{\bf k}\cdot {\bf r}}$ 
with the Bloch function $\psi^{\rm K}_{{\rm c},{\bf k}}$ as 
$\psi^{\rm K}_{{\rm c},{\bf k}}({\bf r}) =
N e^{i{\bf k}\cdot {\bf r}} \psi^{\rm K}_{{\rm c},{\bf k}}$ where
$N$ is a normalization constant satisfying $N^2 S=1$,
$S$ is the area (volume) of the system, and
\begin{align}
 \psi^{\rm K}_{{\rm c},{\bf k}} \equiv \frac{1}{\sqrt{2}}
 \begin{pmatrix}
  1 \cr e^{i\Theta({\bf k})}
 \end{pmatrix}.
\label{eq:wfc}
\end{align}
Here ${\bf k}$ is measured from the K point,
and $\Theta({\bf k})$ is defined by an angle of 
${\bf k}=(k_x,k_y)$ measured from the $k_x$-axis as
$(k_x,k_y) \equiv |{\bf k}| 
(\cos \Theta({\bf k}), \sin \Theta({\bf k}))$.
The eigen value of this state is $E=+v_{\rm F}|{\bf p}|$.
The energy eigenstate with the energy eigen value 
$E=-v_{\rm F}|{\bf p}|$
in the valence energy band is written by
\begin{align}
 \psi^{\rm K}_{v,{\bf k}}({\bf r})
 = \frac{e^{i{\bf k}\cdot {\bf r}}}{\sqrt{2S}}
 \begin{pmatrix}
  1 \cr
  -e^{+i\Theta({\bf k})}
 \end{pmatrix}.
\label{eq:wfv}
\end{align}
The energy eigenstate for the valence band,
$\psi^{\rm K}_{v,{\bf k}}({\bf r})$ is given by 
$\sigma_z \psi^{\rm K}_{c,{\bf k}}({\bf r})$.
This results from the particle-hole symmetry of the Hamiltonian:
$\sigma_z {\cal H}^{\rm K}_0 \sigma_z 
=-{\cal H}_0^{\rm K}$.

%%%%%%%%%%%%%%%%%%%%%%%%%%%%%
%%%%%%%%%%%%%%%%%%%%%%%%%%%%%

The unperturbed Hamiltonian near the K$'$ point
is given by 
\begin{align}
 {\cal H}^{\rm K'}_0 
 = v_{\rm F} \bsigma' \cdot \hat{\mathbf{p}},
 \label{eq:H0K'}
\end{align}
where $\bsigma'=(-\sigma_x,\sigma_y)$.
The dynamics of $\pi$-electrons near the K$'$ point relates to 
the electrons near the K point by time-reversal symmetry,
$\psi^{\rm K} \to (\psi^{\rm K'})^*$ [\cite{sasaki08ptps}].
Because lattice vibrations do not break 
time-reversal symmetry,
we mainly consider electrons near the K point in this paper.

\subsection{Deformation-Induced Gauge Field}\label{subsec:digf}

Lattice deformation modifies
the nearest-neighbor hopping integral locally 
as $-\gamma_0 \to -\gamma_0+\delta \gamma_0^a({\bf r}_i)$ 
($a=1,2,3$) (see Fig.~\ref{fig:graphene}).
The corresponding perturbation of the lattice deformation 
is given by
\begin{align}
 {\cal H}_1 \equiv 
 \sum_{i \in {\rm A}} \sum_{a=1,2,3} 
 \delta \gamma^a_0(\mathbf{r}_i) 
 \left[
 (c_{i+a}^{\rm B})^\dagger c_i^{\rm A} + 
 (c_i^{\rm A})^\dagger c_{i+a}^{\rm B}
 \right],
 \label{eq:H1}
\end{align}
where $c_i^{\rm A}$ is the annihilation operator of 
a $\pi$ electron of an A-atom at position ${\bf r}_i$,
and $(c^{\rm B}_{i+a})^\dagger$ is a creation operator
at position ${\bf r}_{i+a}$ $(={\bf r}_i+{\bf R}_a)$
of a B-atom where ${\bf R}_a$ ($a=1,2,3$) are vectors 
pointing to the three nearest-neighbor B sites from an A site.

The perturbation of Eq.~(\ref{eq:H1})
gives rise to scattering within a 
region near the K point (intravalley
scattering) whose interaction
is given by a deformation-induced gauge field 
${\bf A}({\bf r})=(A_x({\bf r}),A_y({\bf r}))$
in Eq.~(\ref{eq:H0K}) as 
\begin{align}
 {\cal H}^{\rm K}_0 + {\cal H}^{\rm K}_1 = 
 v_{\rm F} \bsigma \cdot \left[ \hat{\mathbf{p}}+{\bf A}^{\rm q}({\bf r}) \right],
 \label{eq:W}
\end{align}
where 
${\bf A}^{\rm q}({\bf r})$ is defined from 
$\delta \gamma_{0,a}({\bf r})$ ($a=1,2,3$) as
[\cite{sasaki05,sasaki06jpsj,katsnelson08}]
\begin{align}
 \begin{split}
  & v_{\rm F} A^{\rm q}_x({\bf r}) = \delta \gamma_{0,1}({\bf r})
  - \frac{1}{2} \left[ \delta \gamma_{0,2}({\bf r}) +
  \delta \gamma_{0,3}({\bf r}) \right], \\
  & v_{\rm F} A^{\rm q}_y({\bf r}) = \frac{\sqrt{3}}{2} 
  \left[ \delta \gamma_{0,2}({\bf r}) -
  \delta \gamma_{0,3}({\bf r}) \right].
 \end{split}
 \label{eq:A}
\end{align}
When $\delta \gamma_{0,2}=\delta \gamma_{0,3}=0$,
then ${\bf A}^{\rm q}({\bf r})=(A_x({\bf r}),0)$ and 
${\bf A}^{\rm q}({\bf r})\cdot {\bf R}_1=0$.
Similarly, when $\delta \gamma_{0,1}=\delta \gamma_{0,3}=0$,
we have ${\bf A}^{\rm q}({\bf r})\cdot {\bf R}_2=0$.
Generally, the direction of ${\bf A}^{\rm q}({\bf r})$
is pointing perpendicular to the bond 
whose hopping integral is changed from $\gamma_0$.
For the K$'$ point, 
we obtain
\begin{align}
 {\cal H}^{\rm K'}_0 + {\cal H}^{\rm K'}_1 = 
 v_{\rm F} \bsigma' \cdot \left[ \hat{\mathbf{p}}-{\bf A}^{\rm q}({\bf r}) \right].
\end{align}
Even though the ${\bf A}^{\rm q}({\bf r})$ appears as a gauge field,
it does not break time-reversal symmetry
because the sign in front of ${\bf A}^{\rm q}({\bf r})$ 
is opposite to each other for the K and K$'$ points.
This is in contrast with the fact that 
${\bf A}({\bf r})$ (vector potential)
violates time-reversal symmetry
because the sign in front of ${\bf A}({\bf r})$ 
is the same for the K and K$'$ points
since ${\hat {\bf p}} \to {\hat {\bf p}}-e{\bf A}({\bf r})$
in the presence of a magnetic field.

The gauge field description
for the lattice deformation (Eq.~(\ref{eq:A})) 
is useful to show the appearance 
of the curvature-induced mini-energy gap  
in metallic carbon nanotubes.
For a zigzag nanotube, 
we have $\delta \gamma_{0,1}=0$ and 
$\delta \gamma_{0,2}=\delta \gamma_{0,3} \ne 0$ from the
rotational symmetry around the tube axis  (see
Fig.~\ref{fig:graphene}). 
Then, Eq.~(\ref{eq:A}) shows that for $A^{\rm q}_x \ne 0$ and $A^{\rm q}_y = 0$, 
the cutting line of $k_x=0$ for the metallic zigzag
nanotube is shifted by a finite constant value of $A^{\rm q}_x$ because of
the Aharanov-Bohm effect for the lattice distortion-induced gauge
field ${\bf A}^{\rm q}$. 
For an armchair nanotube, 
we have $\delta \gamma_{0,1} \ne 0$ and 
$\delta \gamma_{0,2}=\delta \gamma_{0,3}$.
Then, Eq.~(\ref{eq:A}) shows that for $A^{\rm q}_x \ne 0$ and $A^{\rm q}_y = 0$, 
the cutting line of $k_y=0$ for the armchair nanotube 
is not shifted by a vanishing $A^{\rm q}_y$.
This explains the presence (absence) of the curvature-induced
mini-energy gap in metallic zigzag (armchair) carbon nanotubes
[\cite{kane97}].

The gauge field description
is also useful to discuss the effect of an uniaxial strain on the gap.
Let us consider applying a strain along the axis of a zigzag SWNT. 
Then, due to the symmetry, we have $\delta \gamma_{0,1}=a$ and 
$\delta \gamma_{0,2}=\delta \gamma_{0,3} =a/2$ where $a$ is a constant. 
Putting these perturbations into Eq.~(\ref{eq:A}) 
we see that $v_{\rm F}A^{\rm q}_x =a/2$, 
which means that the curvature-induced gap in a zigzag nanotube
can change a little by the strain along the axis.
For an armchair SWNT, instead, we have
$\delta \gamma_{0,1}=0$ and $\delta \gamma_{0,2}=\delta \gamma_{0,3}
=b$, which results in $v_{\rm F} A^{\rm q}_y=0$.
This shows that the absence of the gap in armchair SWNT is robust
against a strain applied along the nanotube axis.

\subsection{Deformation-Induced Gauge Fields for LO and TO modes}
\label{}

\begin{figure}[htbp]
 \begin{center}
  \includegraphics[scale=0.6]{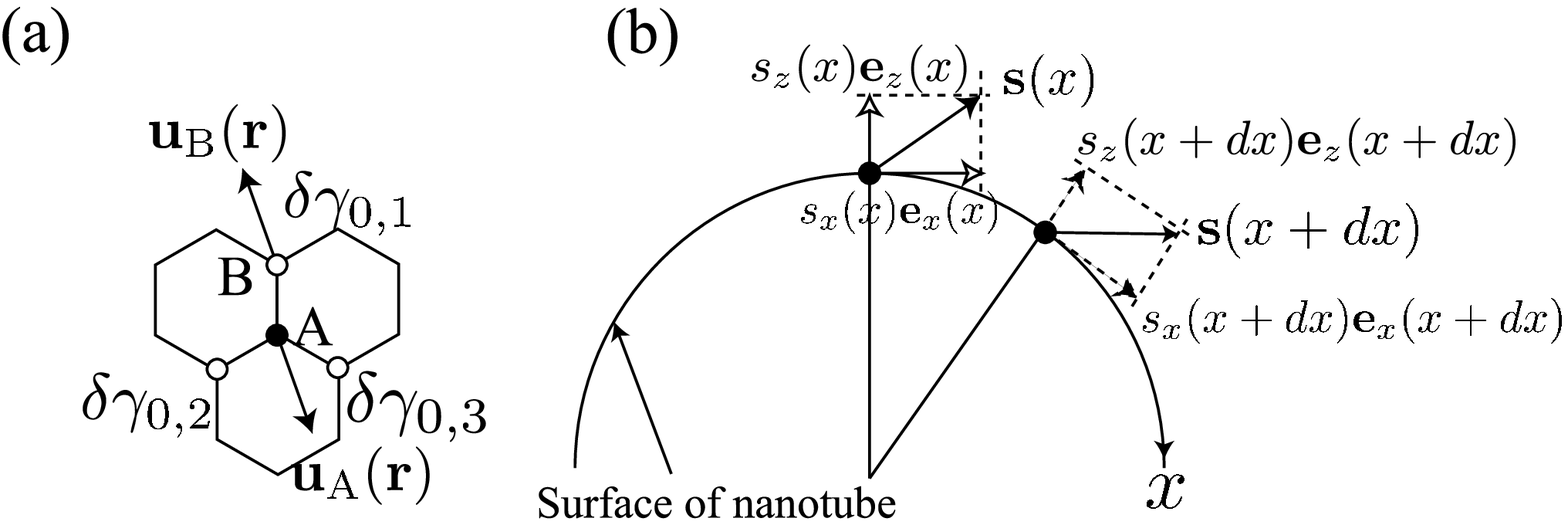}
 \end{center}
 \caption{
 (a) The hexagonal lattice deformed by a phonon displacement vector.
 Let ${\bf u}_{\rm A}({\bf r})$ (${\bf u}_{\rm B}({\bf r})$)
 is a displacement vector of an {\rm A} ({\rm B}) site, 
 then the modulation given by optical phonon modes is 
 $\delta \gamma_{0,a} = (g/\ell) {\bf u}({\bf r}) \cdot {\bf R}_a$
 where ${\bf u}({\bf r})$ 
 ($={\bf u}_{\rm A}({\bf r})-{\bf u}_{\rm B}({\bf r})$) 
 is a relative displacement vector of a {\rm B} site
 relative to the nearest {\rm A} site.
 (b) The cross section of a nanotube.
 The displacement vector for the RBM, ${\bf s}(x)$, 
 is decomposed in terms of the normal $s_z(x)$ and 
 tangential $s_x(x)$ components.
 The derivative of the normal unit vector ${\bf e}_z(x)$ with respect to
 $x$ gives a component along ${\bf e}_x(x)$, which modifies the net
 displacement along the $x$ direction.
 }
 \label{fig:rbm}
\end{figure}

Here, we derive ${\bf A}^{\rm q}({\bf r})$ for the LO and TO modes. 
Let ${\bf u}({\bf r})$ be the relative displacement vector 
of a {\rm B} site from an {\rm A} site 
(${\bf u}({\bf r})={\bf u}_{\rm B}({\bf r})-{\bf u}_{\rm A}({\bf r})$)
and let $g$ be the el-ph coupling constant, then
$\delta \gamma_{0,a}({\bf r})$ for the LO and TO modes is given by
\begin{align}
 \delta \gamma_{0,a}({\bf r})=\frac{g}{\ell} {\bf u}({\bf r}) \cdot {\bf R}_a
\end{align}
where ${\bf R}_a$ denotes the nearest-neighbor vectors 
(Fig.~\ref{fig:graphene} and Fig.~\ref{fig:rbm}(a)) and 
$g=6.4$ eV/\AA\ is the off-site el-ph matrix element [\cite{porezag95}].
We rewrite Eq.~(\ref{eq:A}) as
\begin{align}
 v_{\rm F} (A^{\rm q}_x({\bf r}),A^{\rm q}_y({\bf r})) = g (u_y({\bf r}),-u_x({\bf r})),
 \label{eq:optigauge}
\end{align}
where $u_i({\bf r}) \equiv {\bf u}({\bf r}) \cdot {\bf e}_i$, ($i=x,y$),
and ${\bf R}_1-({\bf R}_2+{\bf R}_3)/2=\ell {\bf e}_y$ 
and $\sqrt{3}/2({\bf R}_2-{\bf R}_3)=-\ell {\bf e}_x$
($\ell \equiv 3a_{\rm cc}/2$)
have been used (see the caption of Fig.~\ref{fig:graphene}).
Then, 
the el-ph interaction 
for an in-plane lattice distortion ${\bf u}({\bf r})$
can be rewritten as the
vector product of $\bsigma$ and
${\bf u}({\bf r})$ [\cite{ishikawa06}] as
\begin{align}
 {\cal H}_{\rm G} \equiv v_{\rm F} \bsigma \cdot {\bf A}^{\rm q}({\bf r})
 =g(\bsigma\times{\bf u}({\bf r}))\cdot {\bf e}_z.
 \label{eq:H_int}
\end{align}

%%%%%%%%%%%%%%%%%%%%%%%%%%%%%
%%%%%%%%%%%%%%%%%%%%%%%%%%%%%

The gauge field description
for the el-ph interaction of the LO and TO modes
(Eq.~(\ref{eq:optigauge})) 
is useful to show the absence of the el-ph interaction for 
the TO mode with a finite wavevector, as shown below.
The TO phonon mode with ${\bf q}\ne 0$ does not change the
area of the hexagonal lattice but instead gives rise to a shear
deformation. Thus, the TO mode (${\bf u}_{\rm TO}({\bf r})$)
satisfies
\begin{align}
 \nabla \cdot {\bf u}_{\rm TO}({\bf r}) = 0, \ \ \ \
 \nabla \times {\bf u}_{\rm TO}({\bf r}) \ne 0.
 \label{eq:TO}
\end{align}
Using Eqs.~(\ref{eq:optigauge}) and (\ref{eq:TO}), 
we see that 
the TO mode does not yield a deformation-induced magnetic field,
\begin{align}
 {\bf B}^{\rm q}({\bf r})
 \equiv \nabla \times {\bf A}^{\rm q}({\bf r}),
\end{align}
but the divergence of ${\bf A}^{\rm q}({\bf r})$ 
instead does not vanish because
\begin{align}
 \begin{split}
  & B^{\rm q}_z({\bf r})= 
  -\frac{g}{v_{\rm F}} \nabla \cdot {\bf u}_{\rm TO}({\bf r})=0,
  \\ 
  & \nabla \cdot {\bf A}^{\rm q}({\bf r})=
  \frac{g}{v_{\rm F}}
  (\nabla \times {\bf u}_{\rm TO}({\bf r}))\cdot {\bf e}_z
  \ne 0.
\end{split}
 \label{eq:grapheneTO}
\end{align}
Thus, we can define a scalar function $\varphi({\bf r})$
which satisfies ${\bf A}^{\rm q}({\bf r})=\nabla \varphi({\bf r})$.
Since we can set ${\bf A}^{\rm q}({\bf r})=0$ in Eq.~(\ref{eq:W})
by selecting the gauge 
as $\psi^{\rm K}({\bf r}) \to \exp( -i\varphi({\bf r})/\hbar )
\psi^{\rm K}({\bf r})$ [\cite{sasaki05}]
and thus
the ${\bf A}^{\rm q}({\bf r})$ in Eq.~(\ref{eq:W})
disappears for the TO mode with ${\bf q}\ne 0$.
This explains why the TO mode with ${\bf q}\ne 0$ 
is completely decoupled from the electrons,
and that only the TO mode with ${\bf q}= 0$
couples with electrons.
This conclusion is valid even when the graphene sheet 
has a static surface deformation.
In this sense, the TO phonon mode at the $\Gamma$-point 
is anomalous since the el-ph interaction for the TO mode
can not be eliminated by a phase of the wavefunction. 
In contrast, 
the LO phonon mode with ${\bf q}\ne 0$
changes the area of the hexagonal lattice
while it does not give rise to a shear deformation.
Thus, the LO mode (${\bf u}_{\rm LO}({\bf r})$) satisfies
\begin{align}
 \nabla \cdot {\bf u}_{\rm LO}({\bf r}) \ne 0, \ \ \ \
 \nabla \times {\bf u}_{\rm LO}({\bf r}) = 0.
 \label{eq:LO}
\end{align}
Using Eqs.~(\ref{eq:optigauge}) and (\ref{eq:LO}), 
we see that the LO mode gives rise to a 
deformation-induced magnetic field since
\begin{align}
 B_z({\bf r}) \ne 0, \ \ \
 \nabla \cdot {\bf A}({\bf r})= 0.
\end{align}
Since a magnetic field changes the energy band structure
of electrons,
the LO mode can couple strongly to the electrons even for ${\bf q}\ne 0$.

\subsection{Deformation-Induced Gauge Field for the RBM}
\label{}

Next, we derive the deformation-induced gauge field
${\bf A}^{\rm q}({\bf r})$ for the RBM.
When the RBM displacement vector of a carbon atom at ${\bf r}$ is
${\bf s}({\bf r})=(s_x({\bf r}),s_y({\bf r}),s_z({\bf r}))$, 
the perturbation to the nearest-neighbor hopping integral 
is given by
\begin{align}
 \delta \gamma_{0,a}({\bf r}) 
 = \frac{g}{\ell}
 {\bf R}_a \cdot
 \{ {\bf s}({\bf r}+{\bf R}_a)- {\bf s}({\bf r}) \},
 \label{eq:delta_gamma}
\end{align}
By expanding ${\bf s}({\bf r}+{\bf R}_a)$ in a Taylor's series around
the displacement ${\bf s}({\bf r})$ as
${\bf s}({\bf r}+{\bf R}_a)={\bf s}({\bf r})+({\bf R}_a\cdot
\nabla) {\bf s}({\bf r}) + \cdots$, we approximate
Eq.~(\ref{eq:delta_gamma}) as
\begin{align}
 \delta \gamma_{0,a}({\bf r}) 
 \simeq \frac{g}{\ell}
 {\bf R}_a \cdot 
 \left\{ ({\bf R}_a \cdot \nabla) {\bf s}({\bf r}) \right\}.
 \label{eq:dg_ac}
\end{align}
Putting ${\bf R}_1=a_{\rm cc}{\bf e}_y$,
${\bf R}_2=-(\sqrt{3}/2)a_{\rm cc}{\bf e}_x -(1/2)a_{\rm cc}{\bf e}_y$,
and ${\bf R}_3=(\sqrt{3}/2)a_{\rm cc}{\bf e}_x -(1/2)a_{\rm cc}{\bf
e}_y$, into the right-hand side of Eq.~(\ref{eq:dg_ac}), 
we obtain the corresponding deformation-induced gauge field
of Eq.~(\ref{eq:A}) as
\begin{align}
 \begin{split}
  & v_{\rm F}A^{\rm q}_x({\bf r}) = \frac{ga_{\rm cc}}{2} \left[
  -\frac{\partial s_x({\bf r})}{\partial x}+
  \frac{\partial s_y({\bf r})}{\partial y} \right], \\
  & v_{\rm F}A^{\rm q}_y({\bf r}) = \frac{ga_{\rm cc}}{2} \left[
  \frac{\partial s_x({\bf r})}{\partial y}+ 
  \frac{\partial s_y({\bf r})}{\partial x} \right].
 \end{split}
 \label{eq:A_ac}
\end{align}
Further,
the displacements of carbon atoms give 
an on-site deformation potential
in which the diagonal Hamiltonian matrix elements are modified 
by the el-ph interaction [\cite{jiang05prb,saito83}]
\begin{align}
 {\cal H}_{\rm on} = 
 g_{\rm on} \sigma_0 
 \left[
 \frac{\partial s_x({\bf r})}{\partial x} + 
 \frac{\partial s_y({\bf r})}{\partial y}
 \right].
 \label{eq:On_ac}
\end{align}
Here, 
$\partial s_x({\bf r})/\partial x + \partial s_y({\bf r})/\partial y$
($=\nabla \cdot {\bf S}({\bf r})$)
represents the change of the area of a graphene sheet [\cite{suzuura02}].
According to the density functional calculation 
by~\cite{porezag95},
we adopt the on-site coupling constant
$g_{\rm on} =17$[eV].

Since Eqs.~(\ref{eq:A_ac}) and (\ref{eq:On_ac})
are proportional to the derivatives of 
$s_x({\bf r})$ and $s_y({\bf r})$,
that is, they are proportional to ${\bf q}$,
where ${\bf q}$ is the phonon wave vector.
Then, 
the el-ph matrix element for the in-plane longitudinal/transverse 
acoustic (LA/TA) phonon modes vanishes 
at the $\Gamma$ point
Namely,
${\bf A}^{\rm q}({\bf r})=0$ 
and ${\cal H}_{\rm on}=0$
in the limit of ${\bf q}=0$.
Among the TA phonon modes,
there is an out-of-plane TA (oTA) phonon mode.
The oTA mode thus shifts carbon atoms on the flat 2D graphene sheet 
in the $z$-direction [see Fig.~\ref{fig:graphene} and Fig.~\ref{fig:rbm}(b)].
The oTA mode of graphene corresponds to the RBM of a nanotube
even though the RBM is not an acoustic phonon mode [\cite{saito98book}].
In the following, we will show that 
the el-ph interaction for the RBM is enhanced 
due to the curvature of the nanotube
as compared with the oTA mode of graphene
since the RBM is a bond-stretching mode 
due to the cylindrical structure of SWNTs.

The displacements of the RBM modify 
the radius of a nanotube as 
$r \to r + s_z({\bf r})$ (see Fig.~\ref{fig:rbm}(b)).
A change of the radius gives rise to two effects 
to the electronic state.
One effect is a shift of the quantized transverse wave vector around the
tube axis. 
The distance between two wave vectors around the tube axis depends on 
the inverse of the radius due to the periodic boundary condition,
and a change of the radius results in a shift of the wavevector.
The other effect is that 
the RBM can change the area on the surface of the nanotube
even at the $\Gamma$ point.
This results in an enhancement of the on-site el-ph interaction.
These two effects are relevant to the fact that the normal vector 
on the surface of a nanotube is pointing in a different direction 
depending on the atom position.
To show this, we take a (zigzag) nanotube as shown in 
Fig.~\ref{fig:rbm}(b).
Let us denote the displacement vectors 
of two carbon atoms at $x$ and $x+dx$
as ${\bf s}(x)$ and ${\bf s}(x+dx)$, then
an effective length for the displacement along the $x$ axis
between the nearest two atoms is given by
\begin{align}
 D_x={\bf e}_x(x+dx) \cdot \left[
 {\bf s}(x+dx)- {\bf s}(x) \right].
 \label{eq:dis}
\end{align}
By decomposing ${\bf s}(x)$
in terms of a normal and a tangential unit vector as
${\bf s}(x)=s_z(x){\bf e}_z(x)+s_x(x){\bf e}_x(x)$
(see Fig.~\ref{fig:graphene}(b)),
we see that Eq.~(\ref{eq:dis}) becomes
\begin{align}
 D_x 
 &= s_x(x+dx) + s_z(x+dx) {\bf e}_x(x+dx) \cdot {\bf e}_z(x+dx) \nn \\
 &- s_x(x) {\bf e}_x(x+dx) \cdot {\bf e}_x(x) - 
 s_z(x){\bf e}_x(x+dx) \cdot {\bf e}_z(x) \nn \\
 & = dx \left\{
 \frac{\partial s_x(x)}{\partial x}+ \frac{s_z(x)}{r} \right\} + \cdots,
 \label{eq:rbm_sz}
\end{align}
where we have used the following equations:
\begin{align}
 \begin{split}
  & {\bf e}_z(x+dx) 
  = {\bf e}_z(x) + \frac{dx}{r} {\bf e}_x(x) + \cdots, \\
  & {\bf e}_x(x+dx) 
  = {\bf e}_x(x) + \frac{dx}{r} {\bf e}_z(x) + \cdots.
 \end{split}
\end{align}
Equation~(\ref{eq:rbm_sz}) shows that
the net displacement along the $x$ axis is modified
by the curvature of the nanotube as 
$\partial_x s_x({\bf r}) \to \partial_x s_x({\bf r})+s_z({\bf r})/r$.
The correction is negligible
for a graphene sheet ($r \to \infty$), but appears 
as an enhancement factor
to the el-ph interaction in SWNTs.

The el-ph interaction for the RBM is included by replacing 
$\partial_x s_x({\bf r})$ with $\partial_x s_x({\bf r})+s_z({\bf r})/r$
in Eqs.~(\ref{eq:A_ac}) and (\ref{eq:On_ac}).
In Eq.~(\ref{eq:A_ac}), we have an additional 
deformation-induced gauge field,
\begin{align}
 v_{\rm F}A^{\rm q}_x({\bf r})=-\frac{ga_{\rm cc}}{2} \frac{s_z({\bf r})}{r},
\end{align}
for the RBM mode which gives rise to a shift of the wavevector around the
tube axis even at ${\bf q}=0$.
In Eq.~(\ref{eq:On_ac}), 
it is shown that the RBM produces an additional on-site deformation
potential of $g_{\rm on} \sigma_0 (s_z({\bf r})/r)$.
Finally, 
we obtain the el-ph interaction 
for the $\Gamma$ point (${\bf q}=0$: ${\bf s}({\bf r})$ is a constant) RBM, 
as
\begin{align}
 {\cal H}_{\rm RBM} 
 = -\frac{ga_{\rm cc}}{2} \frac{s_z}{r}\sigma_x  + 
 g_{\rm on} \frac{s_z}{r} \sigma_0 
 = \frac{2s_z}{d_t}
 \begin{pmatrix}
  g_{\rm on} & -\frac{ga_{\rm cc}}{2} \cr
  -\frac{ga_{\rm cc}}{2} & g_{\rm on}
 \end{pmatrix}.
 \label{eq:Hep}
\end{align}
This representation is for zigzag SWNTs. 
For a general $(n,m)$ SWNT
with a chiral angle $\theta$,
the el-ph interaction for the RBM becomes
\begin{align}
 {\cal H}_{\rm RBM}(\theta) =
 \frac{2s_z}{d_t}
 \begin{pmatrix}
  g_{\rm on} & -\frac{ga_{\rm cc}}{2} e^{+i3\theta} \cr
  -\frac{ga_{\rm cc}}{2} e^{-i3\theta} & g_{\rm on}
 \end{pmatrix}.
 \label{eq:rbm_int}
\end{align}
See \cite{sasaki08_chiral} for more details.

\section{Kohn Anomaly Effect}
\label{sec:main}

Here we consider the el-ph matrix element as a function of the electron
wavevector ${\bf k}$ for the LO and TO phonon modes and the RBM 
with ${\bf q}={\bf 0}$ (i.e., $\Gamma$-point).
The displacement vector with ${\bf q}={\bf 0}$
is expressed by a position independent
${\bf u}=(u_x,u_y)$, by which an electron-hole pair is excited. 
The el-ph interaction with ${\bf q}={\bf 0}$ is relevant to 
phonon-softening phenomena for all three kinds of modes.

\subsection{Matrix Element for Electron-hole Pair Creation}

Let us first consider the case of a zigzag SWNT.
In Fig.~\ref{fig:graphene},
we denote $y$ ($x$) as a coordinate along (around) the axis of a zigzag
SWNT, and $u_y$ ($u_x$) are assigned to the LO (TO) phonon
mode.~\footnote{In case of the $\Gamma$ point phonon, the definition of the
LO and TO is not unique. It seems standard that the LO is taken as the
mode parallel with respect to the tube axis and the TO mode is the one 
perpendicular to the tube axis.} 
Thus, from Eq.~(\ref{eq:optigauge}), we have
\begin{align}
\begin{split}
 & v_{\rm F}{\bf A}^{\rm q}_{\rm LO}=g(u_y,0), \\
 & v_{\rm F}{\bf A}^{\rm q}_{\rm TO}=g(0,-u_x).
\end{split}
\label{eq:opg-zig}
\end{align}
The direction of the gauge field ${\bf A}^{\rm q}({\bf r})$
is perpendicular to the phonon eigenvector ${\bf u}$
and the LO mode shifts the wavevector around the tube axis,
which explains how
the LO mode may induce a dynamical energy band-gap 
in metallic nanotubes [\cite{dubay02}].
Putting Eq.~(\ref{eq:opg-zig}) into Eq.~(\ref{eq:H_int}), we get
\begin{align}
\begin{split}
 & {\cal H}^{\rm zig}_{\rm LO} = v_{\rm F} {\bf A}^{\rm q}_{\rm LO} \cdot \bsigma = gu_y \sigma_x, \\ 
 & {\cal H}^{\rm zig}_{\rm TO} = v_{\rm F} {\bf A}^{\rm q}_{\rm TO} \cdot \bsigma =-gu_x \sigma_y.
\end{split}
\label{eq:Hzig}
\end{align}
The el-ph matrix element $V_{\bf k}$
for the electron-hole pair generation 
is given from Eqs.~(\ref{eq:wfc}), (\ref{eq:wfv}) and (\ref{eq:Hzig}), 
by $(\lambda={\rm LO},{\rm TO})$
\begin{align}
 \langle {\rm eh}({\bf k})|{\cal H}^{\rm zig}_{\rm \lambda}|\omega_\lambda \rangle
 \equiv \int 
 (\psi^{\rm K}_{c,{\bf k}}({\bf r}))^\dagger {\cal H}^{\rm zig}_{\lambda}
 \psi^{\rm K}_{v,{\bf k}}({\bf r}) d^2{\bf r}.
 \label{eq:mat_elph}
\end{align}
By calculating Eq.~(\ref{eq:mat_elph}) 
for the LO mode with $(u_x,u_y) = (0,u)$ and  
for the TO mode with $(u_x,u_y) = (u,0)$,
we get 
\begin{align}
\begin{split}
 & \langle {\rm eh}({\bf k})|{\cal H}^{\rm zig}_{\rm LO}|\omega_{\rm LO} \rangle
 = -igu \sin \Theta({\bf k}), \\
 & \langle {\rm eh}({\bf k})|{\cal H}^{\rm zig}_{\rm TO}|\omega_{\rm TO} \rangle
 = -igu \cos \Theta({\bf k}),
\end{split} 
\end{align}
where $\Theta({\bf k})$ is defined by an angle of 
${\bf k}=(k_x,k_y)$ measured from the $k_x$ axis.

\begin{figure}[htbp]
 \begin{center}
  \includegraphics[scale=0.7]{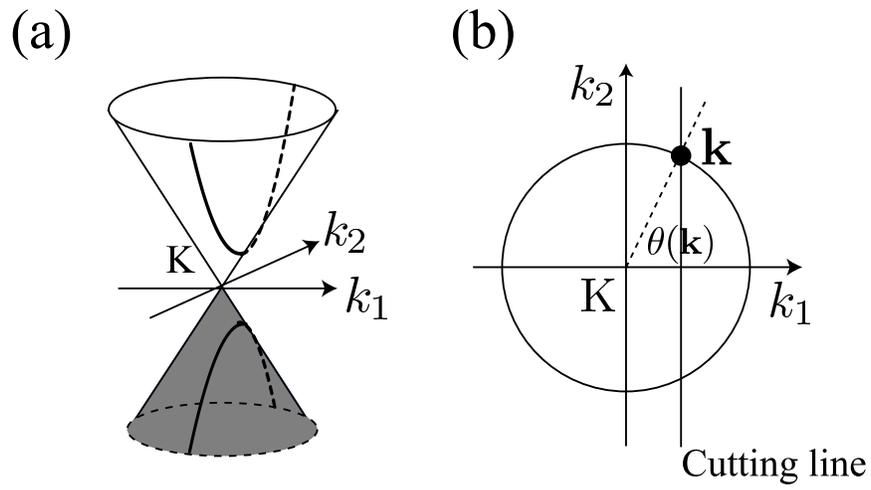}
 \end{center}
 \caption{
 Dependence of LO and TO phonons on the cutting line.
 (a) A cutting line near the K-point.
 The $k_1$ ($k_2$) axis is selected 
 as the nanotube circumferential (axis) direction. 
 The amplitude for an electron-hole pair creation 
 depends strongly on the relative position of the cutting line
 from the K-point.
 (b) If the cutting line crosses the K-point, then
 the angle $\theta({\bf k})$ ($\equiv \arctan(k_2/k_1)$)
 takes $\pi/2$ ($-\pi/2$) values for $k_2 > 0$ ($k_2 < 0$).
 In this case, the LO mode strongly couples to an electron-hole pair,
 while the TO mode is decoupled from the electron-hole pair
 according to Eq.~(\ref{eq:coupling}).
 }
 \label{fig:scattering}
\end{figure}

Next, we consider the case of an armchair SWNT.
In Fig.~\ref{fig:graphene},
$x$ ($y$) is the coordinate along (around) the axis and 
$u_x$ ($u_y$) is assigned to the LO (TO) phonon mode.
Then, for an armchair SWNT, we get
\begin{align}
\begin{split}
 & \langle {\rm eh}({\bf k})|{\cal H}^{\rm arm}_{\rm LO}|\omega_{\rm LO} \rangle
 = -igu \sin \theta({\bf k}), \\
 & \langle {\rm eh}({\bf k})|{\cal H}^{\rm arm}_{\rm TO}|\omega_{\rm TO} \rangle
 = -igu \cos \theta({\bf k}).
\end{split} 
 \label{eq:coupling}
\end{align}
Note that $\theta({\bf k})$ for the armchair nanotube 
is given by rotating $\Theta({\bf k})$ for the zigzag nanotube 
by $\pi/2$ ($\theta({\bf k})=\Theta({\bf k})+\pi/2$).
It is useful to define the $k_1$ ($k_2$) axis 
pointing in the direction of a general SWNT circumferential (axis)
direction (see Fig.~\ref{fig:scattering}), and 
$\theta({\bf k})$ as the angle for the polar coordinate.
Then,
\begin{align}
\begin{split}
 & \langle {\rm eh}({\bf k})|{\cal H}_{\rm LO}|\omega_{\rm LO} \rangle
 = -igu \sin \theta({\bf k}), \\
 & \langle {\rm eh}({\bf k})|{\cal H}_{\rm TO}|\omega_{\rm TO} \rangle
 = -igu \cos \theta({\bf k}).
\end{split} 
 \label{eq:elphG}
\end{align}
is valid regardless of the tube chirality if the 
phonon eigenvector of the LO (TO) phonon mode is 
in the direction along (around) the tube axis.
This is because ${\hat {\bf p}}$ and ${\bf u}({\bf r})$ 
[and ${\bf A}^{\rm q}({\bf r})$]
are transformed in the same way when we change the chiral angle 
[\cite{sasaki08_chiral}].
As a result, 
there would be no chiral angle dependence for the el-ph matrix elements
in Eq.~(\ref{eq:elphG}).
Note also that Eq.~(\ref{eq:elphG}) shows that
$\langle {\rm eh}({\bf k})|{\cal H}_{\lambda}|\omega_\lambda \rangle$
depends only on $\theta({\bf k})$ but not on $|{\bf k}|$,
which means that the dependence 
of this matrix element
on $E^{\rm eh}$ ($=2\hbar v_{\rm F}|{\bf k}|$) is negligible
[see Fig.~\ref{fig:scattering}(b)].

%%%%%%%%%%%%%%%%%%%%%%%%%%%%%
%%%%%%%%%%%%%%%%%%%%%%%%%%%%%

\begin{figure}[htbp]
 \begin{center}
  \includegraphics[scale=0.7]{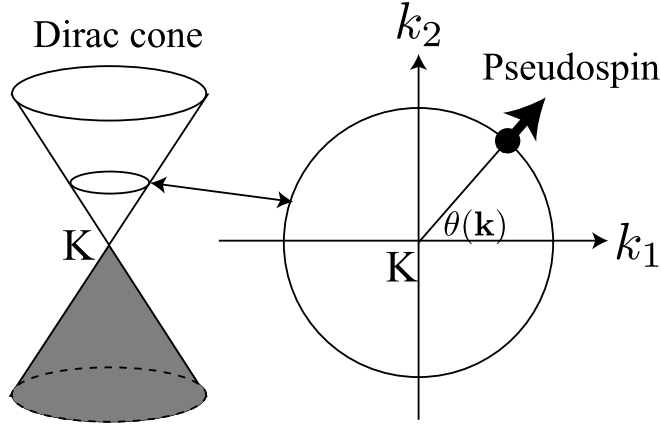}
 \end{center}
 \caption{
 The el-ph matrix element for the electron-hole pair creation process
 relates to the pseudospin of the electronic state.
 }
 \label{fig:zigPhase}
\end{figure}

Where does the $\theta({\bf k})$ dependence in Eq.~(\ref{eq:elphG}) then
come form?
The expectation value of 
$\sigma_x$, $\sigma_y$, and $\sigma_z$
with respect to $\psi^{\rm K}_{{\rm c},{\bf k}}({\bf r})$
defines the pseudospin. 
Using Eq.~(\ref{eq:wfc}) with $\Theta({\bf k})\to \theta({\bf k})$, 
we have the expectation values for the Pauli matrices 
$\langle \sigma_x \rangle =\langle \psi^{\rm K}_{{\rm c},{\bf k}}| \sigma_x |\psi^{\rm K}_{{\rm c},{\bf k}} \rangle
=\cos \theta({\bf k})$,
$\langle \sigma_x \rangle =\langle \psi^{\rm K}_{{\rm c},{\bf k}}| \sigma_y |\psi^{\rm K}_{{\rm c},{\bf k}} \rangle
=\sin \theta({\bf k})$,
and 
$\langle \sigma_x \rangle =\langle \psi^{\rm K}_{{\rm c},{\bf k}}| \sigma_z |\psi^{\rm K}_{{\rm c},{\bf k}} \rangle
=0$.
Then the direction of the pseudospin of $\psi^{\rm K}_{{\rm c},{\bf
k}}({\bf r})$,
\begin{align}
 (\langle \sigma_x \rangle,\langle \sigma_y \rangle, \langle \sigma_z \rangle)=
 (\cos\theta({\bf k}),\sin\theta({\bf k}),0),
\end{align}
is within $(k_1,k_2)$ plane and 
parallel to ${\bf k}$ (see Fig.~\ref{fig:zigPhase}).~\footnote{For the
pseudospin of the electrons near the K$'$ point, see \cite{sasaki10-jpsj}.}
Due to the particle-hole symmetry, 
$\psi^{\rm K}_{v,{\bf k}}({\bf r})=\sigma_z \psi^{\rm K}_{c,{\bf k}}({\bf r})$,
the el-ph matrix element for the electron-hole pair creation process
can be related to the pseudospin.
For example, we see that
\begin{align}
 \langle {\rm eh}({\bf k})|{\cal H}^{\rm zig}_{\rm LO}|\omega_{\rm LO}
 \rangle
 &=gu_y \langle \psi^{\rm K}_{{\rm c},{\bf k}}| \sigma_x |\psi^{\rm
 K}_{{\rm v},{\bf k}} \rangle \nn \\
 &=gu \langle \psi^{\rm K}_{{\rm c},{\bf k}}| \sigma_x \sigma_z
 |\psi^{\rm K}_{{\rm c},{\bf k}} \rangle  \nn \\
 &=-igu \langle \sigma_y \rangle \nn \\
 &=-igu \sin \theta({\bf k}).
\end{align}
The electron-hole pair creation for the LO mode 
is relevant to the pseudospin component which is parallel to
the tube axis, $\langle \sigma_y \rangle$, 
while that for the TO mode is relevant to $\langle \sigma_x \rangle$.

%%%%%%%%%%%%%%%%%%%%%%%%%%%%%
%%%%%%%%%%%%%%%%%%%%%%%%%%%%%

For RBM, from Eq.~(\ref{eq:rbm_int}), 
the matrix element for an electron-hole pair creation 
is chirality dependent as
\begin{align}
 \langle {\rm eh}({\bf k})|{\cal H}_{\rm RBM}(\theta)
 |\omega_{\rm RBM} \rangle =
 iga_{\rm cc}\frac{s_z}{d_t}\sin (\theta({\bf k})+3\theta).
 \label{eq:elph_RBM_g}
\end{align}
Thus, the frequency shift of the RBM 
can have a chiral angle dependence.
In particular, armchair SWNTs ($\theta=30^\circ$)
exhibit neither a frequency shift nor a broadening,
regardless of their diameters
because the el-ph matrix element becomes
\begin{align}
 \langle {\rm eh}({\bf k})|{\cal H}^{\rm arm}_{\rm RBM}
 |\omega_{\rm RBM} \rangle =
 iga_{\rm cc}\frac{s_z}{d_t}\cos \theta({\bf k}),
 \label{eq:aa}
\end{align}
which is zero for a cutting line for a metallic band:
$\theta({\bf k})=\pm \pi/2$.
This $\theta({\bf k})$ dependence of Eq.~(\ref{eq:aa})
is the same as that of the TO phonon mode of Eq.~(\ref{eq:coupling}),
so that the absence of a frequency shift of the RBM in armchair
SWNTs is similar to the absence of a frequency shift of the TO mode at
the $\Gamma$ point in armchair SWNTs [\cite{sasaki08_curvat}].

\subsection{Phonon frequency shift}

Here we show the calculated results for the phonon 
frequency as a function of the Fermi energy.

\subsubsection{Armchair SWNTs}

First, 
we consider Eq.~(\ref{eq:elphG}) 
for a $k$-point (${\bf k}=(k_1,k_2)$)
on the cutting line of an armchair SWNT. 
Since the armchair SWNT is free from the curvature effect,
the cutting line for its metallic energy band 
satisfies $k_1=0$ and lies on the $k_2$ axis.
Thus, we have $\theta({\bf k})=\pi/2$ ($-\pi/2$) for $k_2>0$ ($k_2<0$).
Then, Eq.~(\ref{eq:elphG}) tells us that
only the LO mode can couple to an electron-hole pair
and the TO mode does not couple to an electron-hole pair
for the metallic energy band of an armchair SWNT.
Similarly, Eq.~(\ref{eq:aa}) shows that 
the RBM of an armchair SWNT 
does not show any phonon softening.

\begin{figure}[htbp]
 \begin{center}
  \includegraphics[scale=0.7]{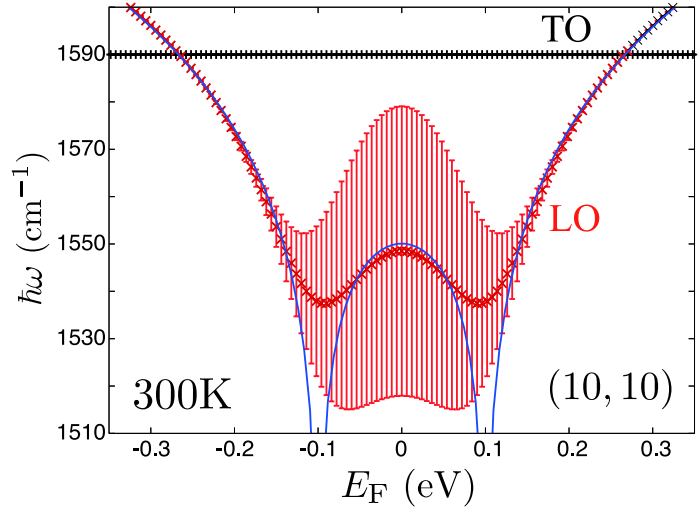}
 \end{center}
 \caption{(color online)
 The $E_{\rm F}$ dependence of the LO (red curve) and TO (black
 curve) phonon energy in the case of the $(10,10)$ armchair SWNT.
 The data are taken at room temperature.
 Only the energy of the LO mode is shifted,
 with the TO mode frequency being independent of $E_{\rm F}$.
 The decay width ($\Gamma$) is plotted as an error-bar.
 The blue curve is given by the analytic result of 
 Eq.~(\ref{eq:selfana}).
 }
 \label{fig:arm}
\end{figure}

In Fig.~\ref{fig:arm},
we show the phonon energy 
as a function of $E_{\rm F}$ for a $(10,10)$ armchair SWNT.
Here we take 1620 ${\rm cm}^{-1}$ and 1590 ${\rm cm}^{-1}$
for $\hbar \omega$ of the LO and TO modes, respectively.
The energy bars denote $\Gamma$ values.
The self-energy is calculated for $T=300$K and $L=10 {\rm \mu m}$.
It is shown that the TO mode does not exhibit any energy change,
while the LO mode shows both an energy shift and a broadening.
As we have mentioned, the minimum energy is realized at 
$|E_{\rm F}| = \hbar \omega/2$ ($\approx 0.1$ eV).
There is a local maximum for the spectral peak 
at $|E_{\rm F}| = 0$.
The broadening for the LO mode 
has a tail at room temperature for $|E_{\rm F}| > \hbar \omega/2$.

%%%%%%%%%%%%%%%%%%%%%%%%%%%%%
%%%%%%%%%%%%%%%%%%%%%%%%%%%%%

In evaluating the LO mode's self-energy according to Eq.~(\ref{eq:PI}),
we have assumed that the cutoff energy is $E_c=0.5$eV.
The presence of a cutoff energy is reasonable since 
the matrix element actually depends on the energy of the electron-hole
pair [see \cite{sasaki09}].
%which is not properly taken into account by the effective-mass model.
An analytical expression for the $E_c$ dependence of the self-energy is
easy to obtain by using the effective-mass model, 
which can be derived from Eq.~(\ref{eq:PI}) at $T=0$ as
\begin{align}
 {\rm Re} \left[ \Pi(\omega,E_{\rm F}) \right] \approx - \frac{L}{\pi}
 \left(\frac{A^{\rm q}_{\rm LO}}{\hbar}\right)^2 \hbar v_{\rm F}
 \left[ 
 \ln \left| \frac{E_c-\frac{\hbar \omega}{2}}{|E_{\rm F}|-\frac{\hbar \omega}{2}} \right|
 +\ln \left| \frac{E_c+\frac{\hbar \omega}{2}}{|E_{\rm F}|+\frac{\hbar \omega}{2}} \right|
 \right].
 \label{eq:selfana}
\end{align}
The factor scales as $1/d_t$ because~\footnote{
Here we use a harmonic oscillator model which gives 
$u=\sqrt{\hbar/2M_c N_u \omega}$ where
$M_c$ is the mass of a carbon atom.
Using $\hbar \omega=0.2$eV, we get
$\sqrt{N_u}|A_x^{\rm q}/\hbar| \approx 2\times 10^{-2}$\AA$^{-1}$.
}
\begin{align}
 \frac{L}{\pi} \left(\frac{A^{\rm q}_{\rm LO}}{\hbar}\right)^2 \hbar v_{\rm F}
 \approx 4[{\rm meV}] \left( \frac{[{\rm nm}]}{d_t} \right).
\end{align}
For comparison, 
we plot $\hbar \omega + {\rm Re} \left[ \Pi(\omega,E_{\rm F}) \right]$
(Eq.~(\ref{eq:selfana})) as the blue curve in Fig.~\ref{fig:arm}.

\subsubsection{Zigzag SWNTs}

Next we consider ``metallic'' zigzag SWNTs.
When the curvature effect is taken into account,
the cutting line does not lie on the K-point, 
but is shifted by $k_1$ from the $k_2$ axis.
In this case,
$\cos \theta({\bf k})=k_1/(k_1^2 + k_2^2)^{1/2}$
is nonzero for the lower energy intermediate 
electron-hole pair states.
Thus, the TO mode can couple to 
the low energy electron-hole pair which
makes a positive energy contribution to the phonon energy shift.
The high energy electron-hole pair 
still decouples from the TO mode
since $\cos\theta({\bf k}) \to 0$ for $|k_2|\gg |k_1|$.

\begin{figure}[htbp]
 \begin{center}
  \includegraphics[scale=0.5]{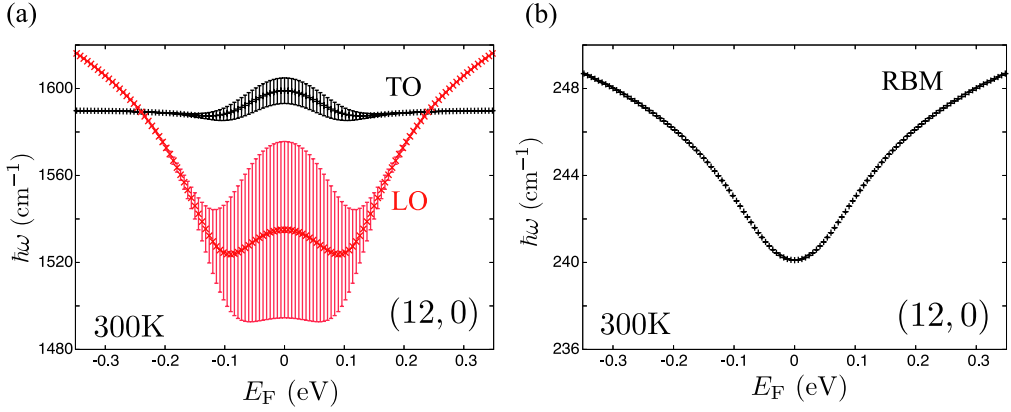}
 \end{center}
 \caption{(color online)
 (a) The $E_{\rm F}$ dependence of the LO (red curve) and TO (black
 curve) phonon energy in the case of the $(12,0)$ zigzag SWNT.
 The data are taken at room temperature. 
 Not only the frequency of the LO mode but also that of the TO mode 
 is shifted due to the curvature effect.
 Here we take 1640 ${\rm cm}^{-1}$ and 1590 ${\rm cm}^{-1}$
 for $\hbar \omega$ of the LO and TO modes, respectively.
 (b) The $E_{\rm F}$ dependence of the RBM energy $\hbar \omega$ 
 in the case of the $(12,0)$ zigzag SWNT.
 }
 \label{fig:zig}
\end{figure}

In Fig.~\ref{fig:zig}(a),
we show calculated results for the LO and TO modes
as a function of $E_{\rm F}$
for a $(12,0)$ zigzag SWNT.
In the case of zigzag SWNTs,
not only the LO mode but also the TO mode
couples with electron-hole pairs.
The spectrum peak position for the TO mode
becomes harder (upshifted) for $E_{\rm F}=0$,
since ${\rm Re}(h_+(E^{\rm eh}))$ for 
$E^{\rm eh} < \hbar \omega$
contributes to a positive frequency shift.
The hardening of the TO mode is a signature 
of the curvature-induced mini-energy gap.

In Fig.~\ref{fig:zig}(b),
we show the result for the RBM.
The matrix element of Eq.~(\ref{eq:elph_RBM_g})
with $\theta=0^\circ$ is proportional to $\sin\theta({\bf k})$.
Thus, the high energy electron-hole pair 
can couple to the RBM and can contribute to the softening of the RBM.
Although the magnitude of the shift is smaller than those for the LO
mode, the softening for the RBM 
can be observed experimentally [\cite{farhat09}].

%%%%%%%%%%%%%%%%%%%%%%%%%%%%%
%%%%%%%%%%%%%%%%%%%%%%%%%%%%%

\subsubsection{Chiral SWNTs}

Finally, we examine ``metallic'' chiral SWNTs.
The same discussion for the ``metallic'' zigzag SWNTs can be applied 
to ``metallic'' chiral SWNTs. 
However, there is a complication specific to chiral SWNTs
that the phonon eigenvector depends on the chiral angle.
\cite{reich01prb} reported that, for a chiral nanotube, 
the atoms vibrate along the direction of the carbon-carbon bonds
and not along the axis or the circumference.
The phonon eigenvector of a chiral nanotube
may be written as
\begin{align}
 \begin{pmatrix}
  u_{\rm TO} \cr u_{\rm LO}
 \end{pmatrix}
 =
 \begin{pmatrix}
  \cos \phi & \sin \phi \cr
  - \sin \phi & \cos \phi
 \end{pmatrix}
 \begin{pmatrix}
  u_1 \cr u_2
 \end{pmatrix},
 \label{eq:chi}
\end{align}
where $u_1$ ($u_2$) is in the direction 
around (along) a chiral tube axis,
and $\phi$ is the angle difference
between the axis and the vibration.
This modifies Eq.~(\ref{eq:coupling}) as
\begin{align}
 \begin{split}
  &\langle {\rm eh}({\bf k})|{\cal H}_{\rm LO}|\omega_{\rm LO} \rangle
  =-ig u \sin (\theta({\bf k})+\phi), \\
  &\langle {\rm eh}({\bf k})|{\cal H}_{\rm TO}|\omega_{\rm TO} \rangle
  =-ig u \cos (\theta({\bf k})+\phi).
 \end{split}
 \label{eq:coupling_chi}
\end{align}
The identification of $\phi$ in Eq.~(\ref{eq:coupling_chi})
as a function of chirality would be useful to compare theoretical
results and experiments, 
which will be explored in the future [See \cite{park09} for example].

\subsubsection{Graphene}

In the case of 2D graphene, 
Eq.~(\ref{eq:coupling}) tells us that 
the $\Gamma$ point TO and LO modes give the same energy shift 
because the integral over $\theta({\bf k})$ gives the same 
self-energy in Eq.~(\ref{eq:PI}) for both TO and LO modes.
This explains why no $G$-band splitting 
is observed in a single layer of graphene [\cite{yan07}].
Even when we consider the TO and LO modes not exactly at the $\Gamma$ point,
we do not expect any splitting between the LO and TO phonon energies
since the TO mode with ${\bf q}\ne 0$
is completely decoupled from the electrons [see
Eq.~(\ref{eq:grapheneTO})].
Thus, for ${\bf q}\ne 0$, only the LO mode contributes to the $G$ band
intensity.
%dependence on $E_{\rm F}$.
It is interesting to note that the $\Gamma$ point LO and TO modes
may exhibit anomalous behavior near the edge of graphene
because the wave function is not given by a plane wave but rather by a
standing wave. 
The pseudospin for the standing wave is different from that
for a plane wave.
Moreover, the standing wave near the zigzag edge is different from that
near the armchair edge, which gives rise to a selection rule 
in their Raman spectra [\cite{sasaki09,sasaki10-jpsj}].
The standing wave behavior near the edges in graphene ribbons 
is beyond the scope of the present paper.

\section{Discussion and Summary}\label{sec:dis}

We have seen that 
the curvature-induced gap is absent for armchair SWNTs, so that
here the LO mode exhibits a strong Kohn anomaly effect. 
Recently, however, it has been reported that 
even armchair SWNTs have an energy gap originating from 
a correlation effect [\cite{deshpande09}].
The correlation-induced gap observed is approximately 80 meV for 
armchair SWNT with $d_t=2$nm, and the gap increases with decreasing $d_t$.
Since the presence of a gap suppresses the contribution to the hardening, 
a local maximum (around $E_{\rm F}=0$) 
in the LO frequency vs. $E_{\rm F}$ plot [see Fig.~\ref{fig:arm}]
may disappear, and $E_{\rm F}=0$ would become a global minimum
if the correlation gap exceeds 200meV.
Confirming this behavior in a Raman spectroscopy study 
may provide further evidence for the correlation-induced gap.

Finally, we discuss our results in relation to 
the experimental results of the Kohn anomaly for the RBMs
in metallic SWNTs [\cite{farhat09}]. 
The frequency shifts observed are approximately 2cm$^{-1}$, 
which is smaller than the theoretical value for a $(12,0)$ zigzag SWNT,
8cm$^{-1}$, shown in Fig.~\ref{fig:zig}(b). 
This discrepancy may be attributed to 
the choice of the off-site el-ph matrix element, $g$, although we
have determined this value from density functional calculation.
Indeed, by decreasing the value of $g$ from $g=6.4$ eV/\AA\
we find a better agreement to the experimental result,
3cm$^{-1}$ when $g=4$ eV/\AA.
Another possibility is that 
we have used a harmonic oscillator model for obtaining the magnitude of
displacement vector $u$. 
Actual value of $u$ may be smaller than our estimation, 
which also gives a better agreement to the experimental result.

In summary, 
the el-ph interaction with respect to the LO, TO, and RBM 
for the Raman feature for SWNTs is derived in a unified way using the
deformation-induced gauge field. 
Then, we have shown that 
the matrix element for electron-hole pair creation 
depends on the position of the cutting line. 
As a result, the TO mode in ``metallic'' SWNTs, 
except for armchair SWNTs,
can couple to an electron-hole pair due to the curvature effect which
shifts the cutting line away from the K point. 
In particular, only the low energy electron-hole pairs 
can couple to the TO mode and give rise to a hardening of the TO mode.
The hardening of the TO mode is suppressed for large diameter SWNTs. 
This is reasonable since the TO mode as well as the LO mode exhibit a
softening in the case of graphene samples.

\section*{Acknowledgment}

K.S. acknowledges a Grant-in-Aid for Specially Promoted Research
(No.~20001006) from MEXT.
R.S acknowledges a MEXT Grant (No.~20241023).
M.S.D acknowledges grant NSF/DMR 07-04197.



%\begin{thebibliography}{00}


%\bibitem{}

%\end{thebibliography}

%\bibliographystyle{apsrev}
\bibliographystyle{elsarticle-harv}
%\bibliographystyle{abbrv}
% \bibliography{/home/sasaki/bib/sasaki_mgm,/home/sasaki/bib/sasaki}

\end{document}